\documentclass[a4paper,11pt]{article}
\usepackage{jheppub}
\usepackage{mathrsfs}
\usepackage{amsfonts}
\usepackage{setspace}
\usepackage{cellspace}
\usepackage{amsmath,bm}
\usepackage[colorlinks=true,linkcolor=blue]{hyperref}
\usepackage{xcolor}
\usepackage{epsfig}
\usepackage{slashed}
\usepackage{caption}
\usepackage{hhline,multirow,tabularx}  % for nicer tables
\usepackage{dcolumn}    % align table columns on decimal point
\usepackage{url}        % for URL addresses
\usepackage{braket}

\title{Spin-spin correlators on the  $\beta$/$\beta^{\star}$  boundaries in 2D Ising-like models: exact analysis through theory of block Toeplitz determinants}
\author[a]{Yizhuang Liu}

\affiliation[a]{Institute of Theoretical Physics,
Jagiellonian University, 30-348 Kraków, Poland}

\emailAdd{yizhuang.liu@uj.edu.pl}

\abstract {In this work, we investigate quantitative properties of correlation functions on the  boundaries between two 2D Ising-like models with dual parameters $\beta$ and $\beta^{\star}$. Spin-spin correlators in such constructions without reflection symmetry with respect to transnational-invariant directions are usually represented as $2\times 2$ block Toeplitz determinants which are usually significantly harder than the scalar ($1\times 1$ block) versions. Nevertheless,  we show that for the specific $\beta/\beta^{\star}$ boundaries considered in this work,  the symbol matrices allow explicit commutative Wiener-Hopf factorizations. As a result, the constants $E(a)$ and $E(\tilde a)$ for the large $n$ asymptotics still allow explicit representations that generalize the strong Szegö's theorem for scalar symbols. However, the Wiener-Hopf factors at different $z$ do not commute. We will show that due to this non-commutativity, ``logarithmic divergences'' in the Wiener-Hopf factors generate certain ``anomalous terms'' in the exponential form factor expansions of the re-scaled correlators. Since our boundaries in the naive scaling limits can be formulated as certain integrable boundaries/defects in 2D massive QFTs, the results of this work facilitate detailed comparisons with bootstrap approaches. }
\date{\today}

\begin{document}
\maketitle
\flushbottom
\section{Introduction}
The exact analysis of spin-spin correlation functions in the bulk of 2D homogeneous Ising models at $h=0$ through theory of Toeplitz determinants~\cite{PhysRev.149.380,PhysRev.155.438,PhysRev.164.719, Wu:1975mw} consolidates the following non-trivial connection: local quantum field theories satisfying the Wightman-Osterwalder-Schrader axioms~\cite{Streater:1989vi,Jost:1965yxu,Osterwalder:1973dx,Osterwalder:1974tc} can be realized through scaling limits of lattice models approaching their critical points~\cite{McCoy:1977er}. Although it is true that before this discovery, there were already several QFTs in 2D and 3D that were rigorously constructed~\cite{Glimm:1987ylb}, these constructions all essentially depend on expansion methods on top of Gaussian free fields. The $h=0$ massive Ising is (are) the first QFT(s) in which non-trivial two-point functions can be fully controlled analytically in a way that does not rely on standard perturbative methods such as Feynman diagram calculations or cluster expansions, yet in the UV limit, reproduces the results of the ``CFT perturbation theory'' organized through power-and-logarithms. 

One must admit that even after almost 50 years, analytical control of two-point functions in massive QFTs to such depth is still very rare. It is possible for the $h=0$ 2D Ising not only due to the large number of conserved charges~\cite{PhysRev.65.117} (e.g., integrability) but also 
due to the fact that the bulk spin-spin correlators are ``dominated by'' Toeplitz determinants with scalar symbols, for which the asymptotic behavior can be analyzed using Wiener-Hopf methods. In particular, due to the fact that the finite $n$ corrections to the spin-spin correlators at $T<T_c$ can be effectively represented as Fredholm determinants with exponentially small Fredholm kernels, the scaling functions can be expressed in the form of {\it exponential form factor expansions} or {\it form factor expansions} which can be generalized to $T>T_c$ as well. It has been made explicit two years later~\cite{cmp/1103901557,cmp/1103904079} after the work~\cite{Wu:1975mw} that the expansion terms in~\cite{Wu:1975mw} can indeed be interpreted physically as form factors: they are (square of) matrix elements of the spin operator between the ground state and ``multi-particle'' states that diagonalize the lattice transfer matrix. In the scaling limit, the multi-particle states become ``asymptotic states'' of a free massive real fermion with trivial scattering, and the scaling form factors satisfy all the ``axioms'' of Lorentz invariance, LSZ reduction, asymptotic completeness, and ``crossing symmetry''. This is not only the first example in, but also inspires the ``form factor bootstrap'' approach~\cite{Smirnov:1992vz} to correlation functions in integrable QFTs with trivial and non-trivial two-particle $S$ matrices. In particular, it has been shown~\cite{Berg:1978sw} that starting from the trivial $S$ matrix $S=-1$ for a massive particle without internal structures, imposing the ``axioms'' for the spin form factors in the ``simplest possible way'' leads exactly to the same scaling function obtained from Toeplitz determinants. 

Given the above, one may ask: if the bootstrap approach is sufficient enough to construct correlation functions in integrable quantum field theories, why formulations of these QFTs through scaling limits of lattice models are still important? To our opinion, this is due to the following reasons. 
Technically, one must admit that most continuum formulations of massive QFTs suffer ambiguities and difficulties: for CFT perturbation theory, one needs to regularize UV and IR divergences, while for the bootstrap approach one also faces uniqueness problems when solving the $S$ matrices and form factors. The standard ``axioms'' in the continuum formulations are also rather artificial for beginners: what is crucial to the analyticity of Wightman functions in the ``permuted region''~\cite{Streater:1989vi,Jost:1965yxu} is the CFT-like short distance asymptotics, which is rather hard to obtain precisely in the bootstrap approach. On the other hand, from the CFT perturbation theory, it is also extremely hard to show the exponential decay at large separations. There always seems to be a wall between the IR and UV asymptotics in the continuum formulations. 

It is only in the lattice formulation that the IR and UV asymptotics of massive QFTs are explained {\it naturally} in prior. Indeed, the lattice formulation is based on the following picture of many near-critical lattice models~\cite{McCoy:1994zi}: ``real physics''  is ``concentrated'' at scales either proportional to the correlation length or to the lattice cutoff, while the ``intermediate regimes'' are saturated by fluctuations without characteristic scales that are universal enough to prevent the transmission of detailed information between the two sides. As such, physics re-scaled to the correlation lengths in such systems are widely conjectured to be universal as well, which in the limit of infinite scale separation,  become exactly massive QFTs that interpolate naturally and smoothly between the algebraic and exponential asymptotics in the UV and IR. 

As such, existence and universality of massive scaling limits (of certain lattice models) should be regarded as among the most fundamental conjectures that are crucial to conceptual understanding of QFTs: not only integrable 2D QFTs, more realistic 4D QFTs with highly nontrivial scatterings (such as QCD) are also conjectured to be realizable through scaling limits of lattice models (such as lattice QCD~\cite{Wilson:1974sk,Wilson:1975id,Osterwalder:1977pc}). Due to this, whenever possible, one should check the agreement between lattice and continuum formulations of the same massive QFT even in the presence of bootstrap constructions, to make sure the following understanding is correct: asymptotics of the specific lattice models in the scaling region are indeed correctly  described by the specific massive integrable QFTs that can also be constructed in the continuum through the bootstrap approach.

In reality, critical lattice models can have boundaries or defects (in 2D it is reasonable to regard boundaries as some form of line defects).  However, the introduction of boundaries or defects also introduces additional discontinuities that break certain spacetime symmetries. As such, UV fluctuations can be potentially enhanced near boundaries and defects. In the moderate case, ``IR physics'' is only modified logarithmically in a universal manner, but in the worst case, there can also be ``power divergences'' that could introduce additional subtleties. This is what happened to Wilson-lines or Wilson-loops in 4D lattice gauge theories due to the combined effects of power divergences and ``renormalon ambiguities''~\cite{Craigie:1980qs,Bauer:2011ws,Bali:2014fea}. In any case, it is reasonable to say that the correlation functions near boundaries or defects are more ``dangerous'' in terms of universality than in the bulk. As such,  comparisons between lattice and continuum formulations must be performed in detail in order understand correctly the scaling asymptotics of defect/boundary correlators in near critical lattice models. 

On the side of continuum formulations of 2D massive QFTs, the notion of integrability has been generalized to include boundaries~\cite{Ghoshal:1993tm} and line defects~\cite{Delfino:1994nr,Bajnok:2009hp} as well. In particular, it has been proposed that near integrable boundaries/defects, particles in the asymptotic states interact with the defects/boundaries {\it elastically} through transmission and reflection coefficients. Compatibility of these coefficients with integrability and other physical requirements introduces severe constraints that have to be solved. Once solved, one can proceed to the calculation of correlation functions either through the ``boundary/defect form factor expansion'' approach treating the direction of the line defect as the Euclidean time~\cite{Bajnok:2006ze,Bajnok:2009hp}, or combining the form factors of the bulk theory with matrix elements of the ``boundary/defect operators'' treating the direction orthogonal to the line defect/boundary as the Euclidean time~\cite{Ghoshal:1993tm,Delfino:1994nr,Konik:1995ws}. For free fermions/bosons, there are many integrable boundaries/defects, but for theories with non-trivial $S$ matrices, integrable boundaries/defects are more restrictive. Nevertheless, partial progress has still been made for the computation of boundary/defect form factors in the Ising, Sinh-Gordon as well as the Lee-Yang models~\cite{Bajnok:2006ze,Bajnok:2009hp}.  

It remains to compare the results of the bootstrap constructions with first principle lattice calculations. For 2D Ising-like models, there are two classical results for boundary/defect propagators. The first is the boundary propagator for the ``half-space'' homogeneous Ising models~\cite{PhysRev.162.436}. Using the Pfaffian method, it has been found that the boundary spin correlators, even in the presence of boundary magnetic fields, allow simple representations through single integrals. Clearly, this is due to the fact that the spin field on the boundary can be chosen to be identical to the free fermion field and consequently, in the form factor expansion one only receives zero-mode and one-particle contributions. Another important result is the spin-spin correlator along a single column/row with modified interactions~\cite{PhysRevLett.44.840}. An important property of this setup is that the lattice remains reflective symmetric with respect to the defect line. As a result, spin-spin correlators along this line can still be represented as Toeplitz determinants with {\it scalar symbols}, for which the multiplicative Wiener-Hopf factorization can be reduced to the additive version for the logarithms. However, without reflection symmetry, correlators along line defects/boundaries in 2D Ising-like models are often represented as two-by-two block determinants~\cite{PhysRevLett.44.840}. Is it still possible to perform exact analysis of the spin-spin correlators along such defects or boundaries and obtain non-trivial form factor expansions?

The purpose of this work is precisely to provide such an example. We will compute the spin-spin correlators on the boundaries between two Ising-like models in 2D with parameters related to each other by the KW-type~\cite{PhysRev.60.252} dualities. One can regard such propagators either as boundary or defect propagators. Crucial to our analysis are the theory of block Toeplitz determinants~\cite{WIDOM1974284,Basor_2007,borodin1999fredholm,basor2000toeplitz} and the fact that the symbol matrices in our cases allow {\it commutative Wiener-Hopf factorization}~\cite{commutative}. To our knowledge, these are the first Ising propagators that have been analyzed exactly based on theory of block Toeplitz determinants and lead to non-trivial form factor expansions in the scaling limit. We will show that the scaling limits of the correlators are rather tricky due to the matrix nature of the Toeplitz symbols. In particular, there are certain terms that we call ``anomalous terms'' that are generated in a subtle manner, against the naive way of taking the scaling limits at the beginning. 
Since our systems in the naive scaling limit can also be formulated as certain integrable line defects for massive free Majorana fermions (see Sec.~\ref{sec:conclude} for more discussions), this example should be of particular interests to be compared with the bootstrap approaches in which the continuum limits are always taken at the very beginning. 

The paper is organized as follows. In Sec.~\ref{sec:summary} we formulate the correlators to be calculated and present the major results of the paper. In Sec.~\ref{sec:matrix}, we investigate in detail the properties of the symbol matrices. We show that they allow commutative Wiener-Hopf factorizations and obtain certain integral representations for certain scalar functions crucial for later use. In Sec.~\ref{sec:III}, armed with the knowledge of matrix Wiener-Hopf, we analyze the correlators in detail. In particular, in subsection.~\ref{sec:E} we calculate the magnetization and their asymptotics near the critical point, and in subsection.~\ref{sec:f} we study the scaling limits of the correlators. We show the presence of ``anomalous terms'' and provide explicit formulas for the exponential form factor expansions up to three-particle form factors. We summarize and make further comments on the work in Sec.~\ref{sec:conclude}.

\section{Formulation of the correlators and summary of results}\label{sec:summary}
We consider the following two systems. First, we introduce the 2D Ising model with the action
\begin{align}\label{eq:Isingaction}
S(\sigma)=&\beta\sum_{k=1}^{M}\sum_{l=1}^{N}\sigma_{k,l}\sigma_{k,l+1}+\beta\sum_{k=0}^{M-1}\sum_{l=1}^{N}\sigma_{k,l}\sigma_{k+1,l}\nonumber \\ 
&+\beta^{\star}\sum_{k=-M}^{-1}\sum_{l=1}^{N}\sigma_{k,l}\sigma_{k,l+1}+\beta^{\star}\sum_{k=-M}^{-1}\sum_{l=1}^{N}\sigma_{k,l}\sigma_{k+1,l} \nonumber \\ 
&+\frac{\beta+\beta^{\star}}{2}\sum_{l=1}^{N}\sigma_{0,l}\sigma_{0,l+1} \ .
\end{align}
Here $e^{-2\beta}=\tanh \beta^{\star}$. The periodic boundary condition is imposed on the horizontal direction ($l$) while the open boundary condition is imposed on the vertical direction ($k$). We are interested in the spin-spin correlator on the $\beta/\beta^{\star}$ boundary
\begin{align}\label{eq:defcorre}
\langle \sigma_{00}\sigma_{0n} \rangle_{\beta}=\lim_{N,M\rightarrow \infty}\frac{\sum_{\{\sigma\} \in \{-1,1\}^{N(2M+1)}}e^{S(\sigma)}\sigma_{00}\sigma_{0n}}{\sum_{\{\sigma\} \in \{-1,1\}^{N(2M+1)}}e^{S(\sigma)}} \ .
\end{align}
Introducing the transfer matrix acting on $\otimes_{l=1}^N R^2$
\begin{align}
\hat T(\beta)=e^{\frac{\beta}{2}\sum_{l=1}^N \sigma^x_l\sigma^x_{l+1}}e^{-\beta^{\star}\sum_{l=1}^N\sigma^z_l}e^{\frac{\beta}{2}\sum_{l=1}^N \sigma^x_l\sigma^x_{l+1}} \ ,
\end{align}
and denote its charge-even ``ground state'' with $(-1)^{\sum_{l=1}^N\sigma_l^+\sigma_l^-}=1$ as $|\Omega_\beta,N\rangle_+$, the correlator can be re-expressed as the following ``quantum'' average
\begin{align}\label{eq:Isingqua}
\langle \sigma_{00}\sigma_{0n} \rangle_\beta=\lim_{N\rightarrow \infty}\frac{\,_+\langle\Omega_{\beta},N|\sigma^x_0\sigma^x_n|\Omega_{\beta^{\star}},N\rangle_+}{\,_+\langle\Omega_{\beta},N|\Omega_{\beta^{\star}},N\rangle_+} \ .
\end{align}
The second system we consider is the ``transverse field Ising chain'' with the Hamiltonian
\begin{align}
\hat H(H)=-\sum_{l=1}^N \sigma^x_l\sigma^x_{l+1}+H\sum_{l=1}^N\sigma_l^z \ ,
\end{align}
with the periodic boundary condition. Again denote the ``charge even'' ground state as $|\Omega_H,N\rangle_+$, one has the similar correlator
\begin{align}\label{eq:correchain}
\langle \sigma_{00}\sigma_{0n} \rangle_H=\lim_{N\rightarrow \infty}\frac{\,_+\langle\Omega_H,N|\sigma^x_0\sigma^x_n|\Omega_{1/H},N\rangle_+}{\,_+\langle\Omega_{H},N|\Omega_{1/H},N\rangle_+} \ .
\end{align}
Without losing generality, we chose $0<H<1$ and $\beta>\beta^{\star}>0$. 

It is not hard to show that these correlators are all given by $2\times 2$ block Toeplitz determinants. For the Ising chain correlator in Eq.~(\ref{eq:correchain}), one introduces the 
$2\times 2$ matrix $a(z)$ with
\begin{align}
& a_{11}(z)=a_{22}(z)=\frac{1-\alpha}{1+\alpha}\frac{1+z}{1-z} \ , \label{eq:dia}\\ 
& a_{12}(z)=\frac{2\sqrt{1-\alpha z}\sqrt{1-\alpha z^{-1}}}{(1+\alpha)(1-z)} \ , a_{21}(z)=za_{12}(z) \ , \label{eq:offdia}
\end{align}
where $0<\alpha=H<1$. For the Ising model correlator, one needs the matrix $\tilde a(z)$ for which $\tilde a_{11}(z)=a_{11}(z)$ and $\tilde a_{22}(z)=a_{22}(z)$ remain the same with the identification $\alpha=e^{-2(\beta-\beta^{\star})}<1$, but the $\tilde a_{12}$ and $\tilde a_{21}$ require the following modifications
\begin{align}
&\tilde a_{12}(z)=\frac{2\sqrt{1-\alpha z}\sqrt{1-\alpha z^{-1}}}{(1+\alpha)(1-z)} \sqrt{\frac{1-\alpha_1z}{1-\alpha_1z^{-1}}} \ , \\ 
&\tilde a_{21}(z)=\frac{2z\sqrt{1-\alpha z}\sqrt{1-\alpha z^{-1}}}{(1+\alpha)(1-z)} \sqrt{\frac{1-\alpha_1z^{-1}}{1-\alpha_1z}} \ ,
\end{align}
where $\alpha_1=e^{-2(\beta+\beta^{\star})}<\alpha<1$. As expected, the Toeplitz symbol for the 2D Ising model is slightly more complicated. 

Naively, in the ``massive scaling limit'' $\alpha \rightarrow 1^-$ with $r=n(1-\alpha)$ fixed~\cite{Wu:1975mw}, one expects that the scaling function, if exists, should be controlled by the behavior of the Toeplitz symbols near $z=1$. Since $\alpha_1$ remains far away from $z=1$ even at $\beta=\beta^{\star}$, one expects that the additional square roots involving $\alpha_1$ should play no role in the ``scaling function''.  Moreover, since $z\approx 1$ in the scaling region, one may even hope that the $z$ factor in the relationship $a_{21}=za_{21}$ can also be approximated by $1$ at the very beginning.  It is the purpose of this work to perform a thorough investigation of the correlators Eq.~(\ref{eq:defcorre}) and Eq.~(\ref{eq:correchain}), especially in the scaling region. We will show that the second hope is incorrect, while the first hope is consistent with our results but in a rather non-trivial manner.  More precisely, the major results of this work are summarized as: 
\begin{enumerate}
    \item In the large $n$ limit, one has 
    \begin{align}
    \langle \sigma_{00}\sigma_{0n}\rangle^2_H \rightarrow E(a) \ne 0 \ , \\
    \langle \sigma_{00}\sigma_{0n}\rangle^2_\beta \rightarrow E(\tilde a) \ne 0 \ .
\end{align}
The constants $E(a)$ and $E(\tilde a)$ scale as $\sqrt{1-\alpha}$ as $\alpha \rightarrow 1^-$:
\begin{align}
& E(a)\rightarrow \sqrt{\frac{1-\alpha}{2}}e^{-\frac{7\zeta_3}{2\pi^2}} \ ,  \label{eq:Eafinal} \\
& E(\tilde a)\rightarrow \sqrt{\frac{1+\alpha_1}{1-\alpha_1}}\sqrt{\frac{1-\alpha}{2}}e^{-\frac{7\zeta_3}{2\pi^2}}  \ .\label{eq:Eatildefinal}
\end{align}
In particular, the ratio $\lim_{\alpha \rightarrow 1^-}\frac{E(a)}{E(\tilde a)}$ remains the same as the homogeneous models.
\item The scaling functions
\begin{align}
F^2_H(r) \equiv \lim_{\alpha \rightarrow 1^{-}}\frac{\langle \sigma_{00}\sigma_{0n}\rangle^2_H}{E(a)}\bigg|_{n=r(1-\alpha)^{-1}} \ ,
\end{align}
and
\begin{align}
F^2_\beta(r) \equiv \lim_{\alpha \rightarrow 1^{-}}\frac{\langle \sigma_{00}\sigma_{0n}\rangle^2_\beta}{E(\tilde a)}\bigg|_{n=r(1-\alpha)^{-1}} \ ,
\end{align}
exit and equal to each other, at least to the second order in the exponential form factor expansions. In particular, their leading large $r$ asymptotics are
\begin{align}
&F_H^2(r)-1 \rightarrow \frac{1}{\sqrt{2\pi}}\frac {e^{-r}}{r^{\frac{3}{2}}}\bigg(e^{-\frac{4G}{\pi}}+{\color{red}e^{\frac{4G}{\pi}}}\bigg)\bigg(1+{\cal O} \left(\frac{1}{r}\right)\bigg) \ , \label{eq:FHlarge}\\ 
&F_\beta^2(r)-1 \rightarrow \frac{1}{\sqrt{2\pi}}\frac {e^{-r}}{r^{\frac{3}{2}}}\bigg(e^{-\frac{4G}{\pi}}+{\color{red}e^{\frac{4G}{\pi}}}\bigg)\bigg(1+{\cal O} \left(\frac{1}{r}\right)\bigg) \ . \label{eq:Fblarge}
\end{align}
Here $G=\sum_{k=0}^{\infty}\frac{(-1)^k}{(2k+1)^2}$ is the Catalan's constant. 

\item 
Beyond the leading large $r$ asymptotics, the scaling functions allow exponential form factor expansions of the following forms 
\begin{align}
    F_H^2(r)=\exp \bigg(\sum_{k=1}^{\infty}\frac{1}{\pi^k}f_k(r)\bigg) \ , \\ 
    F_\beta^2(r)=\exp \bigg(\sum_{k=1}^{\infty}\frac{1}{\pi^k}\tilde f_k(r)\bigg) \ ,
\end{align}
where $f_k(r),\tilde f_k(r)={\cal O}(e^{-kr})$ in the large $r$ limits. Up to $k=3$, $f_{k}(r)=\tilde f_{k}(r)$ and their explicit forms are given here as 
\begin{align}
&f_1(r)=\int_1^{\infty} dt\frac{\sqrt{t^2-1} e^{-tr}}{t^2}\bigg(\frac{t}{g(t)}+{\color{red} \frac{g(t)}{t}}\bigg) \ , \label{eq:f1} \\
&f_2(r)+\frac{1}{2}f_1^2(r) \nonumber \\ 
&=\int_1^{\infty}\int_1^{\infty} dt_1dt_2 \frac{\sqrt{(t_1^2-1)(t_2^2-1)}e^{-t_{12}r}}{t_1t_2t_{12}^2}\bigg(\frac{t_1^2}{g(t_1)g(t_2)}+\frac{g(t_1)g(t_2)}{t_1^2}+{\color{red}\frac{t^2_{12} g(t_1)}{t_1^2g(t_2)}}\bigg) \ , \label{eq:f2} \\
&f_3(r)-\frac{1}{3}f_1^3(r)=-\int_1^{\infty}\int_1^{\infty}\int_1^{\infty} dt_1dt_2dt_3 \frac{\sqrt{(t_1^2-1)(t_2^2-1)(t_3^2-1)}e^{-t_{123}r}}{t_1^3t_3^3t_{12}t_2t_{23}} \nonumber \\ 
&\times \frac{t_1^3t_3^3+{\color{red}g^2(t_1)g^2(t_2)g^2(t_3)+g^2(t_1)g^2(t_3)t_{12}t_{23}+g^2(t_1)t_3^3t_{12}+g^2(t_3)t_1^3t_{23}}}{g(t_1)g(t_2)g(t_3)} \ , \label{eq:f3}
\end{align}
where $t_{12}=t_1+t_2$, $t_{23}=t_2+t_3$, $t_{123}=t_1+t_2+t_3$, and 
\begin{align}
g(t)=(1+t)\exp \bigg(\frac{2}{\pi}\int_{1}^{\infty} du \frac{\arctan \frac{1}{\sqrt{u^2-1}}}{t+u}\bigg)\ .
\end{align}
Notice that in the large $t$ limit one has $g(t)=(1+t)\left(1+{\cal O}\left(\frac{\ln t}{t}\right)\right)$, and consequently the small $r$ asymptotics of $f_1(r)$, $f_2(r)$ and $f_3(r)$ are free from ``power-divergences'' of the form $\frac{1}{r^l}$ with $l>0$. 

\item The terms in the Eq.~(\ref{eq:FHlarge}), Eq.~(\ref{eq:Fblarge}) and the Eq.~(\ref{eq:f1}), Eq.~(\ref{eq:f2}), Eq.~(\ref{eq:f3}) shown in red are generated in a rather subtle manner through some form of  ``anomaly-mechanism'' caused by non-commutativity of Wiener-Hopf factors that amplifies the would be ``power-corrections'' in $z-1$ in the scaling region. In particular, this shows that for $a_{21}=za_{12}$, one can not set $z=1$ at the beginning and proceed with $a_{21}=a_{12}$. On the other hand, although the scaling functions in the two formulations are still the same at least to the order we reached, this is also realized in a rather subtle manner. It requires very specific relations between the non-universal constants in the Wiener-Hopf factors and the coefficients of the anomalous terms.  
\end{enumerate}

\section{Block Toeplitz determinants and their commutative Wiener-Hopf factorization}\label{sec:matrix}
More precisely, due to the fact that the ground states $|\Omega_H,N\rangle_+$ and $|\Omega_{1/H},N\rangle_+$ in the Eq.~(\ref{eq:correchain}) are all ``free'' in the sense that their wave functions in the fermionic coherent states are all exponential functions of quadratic forms, the Eq.~(\ref{eq:correchain}) can still be calculated as a Pfaffian in terms of the ``fermionic two point functions''. Straightforward calculations lead to
\begin{align}\label{eq:corrrelator}
\langle \sigma_{00}\sigma_{0n} \rangle_H^2=D_n(\hat a)={\rm det}_{2n\times 2n} \ T_n(\hat a) \ ,
\end{align}
where $T_n(\hat a)=P_nT(\hat a)P_n$ is the semi-infinite block Toeplitz operator $\hat a_{ij}|_{i, j\ge 0}\equiv \hat a_{i-j}|_{i, j\ge 0}$ projected to the upper-left $n\times n$ block entries with the restriction $0\le i,j \le n-1$.  The element $\hat a_{i-j}$ is defined as 
\begin{align}
\hat a_{i-j}=\frac{1}{2\pi i}{\rm PV}\oint_{C_1} \frac{dz}{z} z^{i-j}a(z) \ , 
\end{align}
where $C_\eta$ denotes the circle with radius $\eta$, and one has the principal value prescription for the pole at $z=1$. Throughout this paper, the directions for the $\oint_{C_\eta}$ are always counter-clockwise.  Given the principal value prescription, the $T_n(\hat a)$ is actually anti-symmetric. This is manifest since $\frac{1+z}{1-z}$ is anti-symmetric under $\theta \rightarrow -\theta$ (we use $z=e^{i\theta}$), while 
\begin{align}
\hat a_{j-i;12}&=\frac{1}{2\pi}{\rm PV}\int_{-\pi}^{\pi}d\theta\frac{2\sqrt{(1-\alpha e^{i\theta})(1-\alpha e^{-i\theta})}}{(1+\alpha)(1-e^{i\theta})}e^{i(j-i)\theta}\nonumber  \\
&=-\frac{1}{2\pi}{\rm PV}\int_{-\pi}^{\pi}d\theta\frac{2e^{i\theta}\sqrt{(1-\alpha e^{i\theta})(1-\alpha e^{-i\theta})}}{(1+\alpha)(1-e^{i\theta})}e^{i(i-j)\theta}\equiv -\hat a_{i-j;21} \ . 
\end{align}
Notice that for $\alpha=1$, the $a_{11}$ and $a_{22}$ all vanish, and the block determinant factorizes into a product of two identical Toeplitz determinants for the homogeneous model at the critical parameter.

For $\alpha\ne 1$, the presence of the principal value is not convenient for the following analysis. To facilitate the analysis, one introduces the matrix
\begin{align}\label{eq:defaij}
a_{i-j}=\frac{1}{2\pi i}\oint_{C_\eta} \frac{dz}{z} z^{i-j}a(z) \equiv \frac{\eta^{i-j}}{2\pi}\int_{-\pi}^{\pi}d\theta e^{i\theta(i-j)}a(\eta e^{i\theta})\ ,
\end{align}
with the integration path chosen to be along a circle $C_\eta$ with radius $\alpha<\eta<1$. Clearly, the matrix $a(z)$ is analytic and has determinant $1$ within this region.   The motivation of introducing the block matrix $a_{i-j}$ is,  for all $i-j \in Z$, one can write 
\begin{align}
a_{i-j}=\hat a_{i-j}+\frac{1-\alpha}{1+\alpha}vv^T \ , 
\end{align}
with $v^T=(1,1)$ is a constant vector in $R^2$. Now, introducing the vector $v_n=\otimes_{k=1}^n v$, one has
\begin{align}
    T_n(a)=T_n(\hat a)+\frac{1-\alpha}{1+\alpha}v_nv_n^T \ . 
\end{align}
The point is, if $T_n(\hat a)$ is invertible, then due to the antisymetry of $T_n(\hat a)$, it is easy to show that 
\begin{align}\label{eq:equality}
D_n(\hat a)={\rm det}_{2n\times 2n} \ T_n(\hat a)=D_n(a)={\rm det}_{2n\times 2n} \  T_n(a) \ .
\end{align}
On the other hand, if $\hat T_n(a)$ is not invertible, then its rank can at most be $2n-2$ and adding an operator with rank $1$ will never make it invertible. Given the above, 
Eq.~(\ref{eq:equality}) is always true, and the task of calculating $D_n$ then reduces to the block determinant with symbol $a_{i-j}$. Notice that the construction above holds for the 2D Ising model with the symbol matrix $\tilde{a}$ as well. In particular, one has
\begin{align}
\langle \sigma_{00}\sigma_{0n} \rangle^2_{\beta}=D_n(\tilde a)={\rm det}_{2n\times 2n} \ T_n(\tilde a) \ , 
\end{align}
where $D_n(\tilde a)$ is defined in the same way as  Eq.~(\ref{eq:defaij}) with $a(z)$ replaced by the $\tilde a(z)$.

Now, we introduce the {\it first-order} polynomial matrix for the Ising chain
\begin{align}
&J(z)=\bigg(\begin{array}{cc}
      0 \  \ 1\\ 
     z  \ \  0
\end{array}\bigg) \ ,
\end{align}
and for the Ising model
\begin{align}
&\tilde J(z)=\frac{1}{1-\alpha_1}\bigg(\begin{array}{cc}
      0 \  \ 1-\alpha_1z\\ 
     z-\alpha_1  \ \  0
\end{array}\bigg) \ .
\end{align}
The crucial fact is , the matrices $a$ and $\tilde a$ allow the exponentiation 
\begin{align}
&a=\exp \bigg(J(z)\times \frac{1}{\sqrt{z}}{\rm arctanh} \frac{2\sqrt{z}\sqrt{(1-\alpha z)(1-\alpha z^{-1})}}{(1-\alpha)(z+1)}  \bigg) \ , \\ 
&\tilde a=\exp \bigg(\tilde J(z)\times \frac{1-\alpha_1}{\sqrt{(1-\alpha_1z)(1-\alpha_1z^{-1})}}\frac{1}{\sqrt{z}}{\rm arctanh} \frac{2\sqrt{z}\sqrt{(1-\alpha z)(1-\alpha z^{-1})}}{(1-\alpha)(z+1)}  \bigg) \ , 
\end{align}
where $z=|z|e^{i\theta}$ with $-\pi<\theta<\pi$, $\sqrt{z}=\sqrt{|z|}e^{\frac{i\theta}{2}}$ and the logarithm in the ${\rm arctanh}$ is defined with the principal branch. Notice that although there is a $\sqrt{z}$ in the definition, the functions
\begin{align}
&f(z,\alpha)=\frac{1}{\sqrt{z}}{\rm arctanh} \frac{2\sqrt{z}\sqrt{(1-\alpha z)(1-\alpha z^{-1})}}{(1-\alpha)(z+1)} \ , \\ 
&\tilde f(z,\alpha,\alpha_1)=\frac{1-\alpha_1}{\sqrt{(1-\alpha_1z)(1-\alpha_1z^{-1})}}f(z,\alpha) \ ,
\end{align}
are in fact single-valued analytic functions in the annulus region $\alpha<|z|<1$. Moreover, $f(z,\alpha)$ can be analytically continued outside the annulus with branch cut singularities along the real axis in $(1,\infty)$ and $(0,\alpha)$. The simplest way to see this is to use the representation in $\alpha<|z|<1$
\begin{align}\label{eq:falter}
f(z,\alpha)=&\frac{1}{\sqrt{z}}\ln \left(1+\frac{(1-\alpha)(z+1)}{2\sqrt{z}\sqrt{(1-\alpha z)(1-\alpha z^{-1})}}\right)+\frac{1}{\sqrt{z}}\ln \frac{2\sqrt{z}\sqrt{(1-\alpha z)(1-\alpha z^{-1})}}{(1-\alpha)(z+1)} \nonumber \\ 
-&\frac{1}{\sqrt{z}}\ln \frac{(1+\alpha)(1-z)}{(1-\alpha)(1+z)} \ ,
\end{align}
from which the analytic continuation to the outside can be performed with the principal branch for all the logarithms and square roots. In particular, there is no singularity in the negative real axis.  

Since the analyticity structure of $f(z,\alpha)$ is crucial, here we provide more explanations on how it can be obtained. We need to show the absence of singularities away from the real axis and then the absence of singularities in the negative real axis. To show the former, we only need to notice that with the principal branch for all the square roots, one has
\begin{align}
{\rm Re} \bigg(\frac{(z+1)}{\sqrt{z}\sqrt{(1-\alpha z)(1-\alpha z^{-1})}}\bigg) \ge 0 \ ,
\end{align}
for all $z\ne 0$ and $0<\alpha<1$. This implies that the arguments of the first two logarithms will never cross the negative real axis, thus the singularities can only due to branch cuts and zeros of the arguments of the logarithms and the square roots,  which are all located in the real axis. To show the absence of singularities in the negative real axis, first calculate $f(-r+i0)-f(-r-i0)$ separately for $0<r<1$ and $r>1$ and notice that they all vanish. Then use the fact that $f$ is bounded around $z=-1$ to show the absence of singularity at $z=-1$ as well. Given the above, singularities of $f$ can only be along the non-negative real axis. From the $\sqrt{(1-\alpha z)(1-\alpha z^{-1})}$ in the first two logarithms, one obtains branch cuts along $(0,\alpha)$ and $(\alpha^{-1},\infty)\subsetneq (1,\infty)$, while from the last logarithm there is another branch cut along $(1,\infty)$. Also notice that around $z=0$, $|f|$ scales as $\frac{1}{\sqrt{|z|}}$ but not $\frac{1}{|z|}$.

The above singularity structure immediately implies the existence of the additive Wiener-Hopf factorization for $\alpha<\eta<1$
\begin{align}
&f^{\pm}(z,\alpha)=\frac{\mp }{2\pi i}\oint_{C_\eta} dz' \frac{f(z',\alpha)}{z-z'} \ , \\ 
&\tilde f^{\pm}(z,\alpha,\alpha_1)=\frac{\mp }{2\pi i}\oint_{C_\eta} dz' \frac{\tilde f(z',\alpha,\alpha_1)}{z-z'} \ ,
\end{align}
which are actually $\eta$-independent and analytic respectively in the regions $|z|<1$ (for $f^+$, $\tilde f^+$) and $|z|>\alpha$ (for $f^-$, $\tilde f^-$).  The equalities
\begin{align}
&f(z,\alpha)=f^+(z,\alpha)+f^-(z,\alpha) \ , \\ 
&\tilde f(z,\alpha,\alpha_1)=\tilde f^+(z,\alpha,\alpha_1)+\tilde f^-(z,\alpha,\alpha_1) \ ,
\end{align}
hold within the region $\alpha<|z|<1$. Also notice that $f^-$ and $\tilde f^-$ decay at infinity at the speed $\frac{1}{z}$. Moreover, by picking up the branch discontinuities using Eq.~(\ref{eq:falter}),  for generic $0<\alpha<1$ one obtains the following explicit representations for the moments $f_k=\oint_{C_\eta} \frac{dz}{2\pi i z}z^kf(z,\alpha)$, $k\in Z$
\begin{align}
&f_{-k}=\frac{1}{\pi}\int_{0}^{\alpha} dx x^{k-\frac{1}{2}}\arctan\left(\frac{(1-\alpha)(1+x)}{2\sqrt{(1-\alpha x)(\alpha-x)}}\right)+\frac{2-\alpha^{k+\frac{1}{2}}}{2k+1} \ ,  \ k \ge 0 \ , \label{eq:fkminus}\\
&f_{k+1}=\frac{1}{\pi}\int_{0}^{\alpha} dx x^{k-\frac{1}{2}}\arctan\left(\frac{(1-\alpha)(1+x)}{2\sqrt{(1-\alpha x)(\alpha-x)}}\right)-\frac{\alpha^{k+\frac{1}{2}}}{2k+1} \ ,\  k \ge 0 \ . \label{eq:fkplus}
\end{align}
Given the above, one has 
\begin{align} \label{eq:fplus}
&f^{+}(z,\alpha)=\sum_{k=0}^{\infty} z^kf_{-k}\nonumber \\ 
&=\frac{1}{\pi}\int_{0}^{\alpha}\frac{dx}{\sqrt{x}(1-zx)}\arctan \frac{(1-\alpha)(1+x)}{2\sqrt{(1-\alpha x)(\alpha-x)}}+\sum_{k=0}^{\infty}\frac{2-\alpha^{k+\frac{1}{2}}}{2k+1}z^k \ ,
\end{align}
and
\begin{align}\label{eq:fminus}
&f^{-}(z,\alpha)=\sum_{k=0}^{\infty} \frac{f_{k+1}}{z^{k+1}}\nonumber \\ 
&=\frac{1}{\pi}\int_{0}^{\alpha}\frac{dx}{\sqrt{x}(z-x)}\arctan \frac{(1-\alpha)(1+x)}{2\sqrt{(1-\alpha x)(\alpha-x)}}-\sum_{k=0}^{\infty}\frac{\alpha^{k+\frac{1}{2}}}{2k+1}\frac{1}{z^{k+1}} \ .
\end{align}
From the above, it is clear that $f^+$ has the natural analyticity region $|z|<1$, while $f^-$ has the natural analyticity region $|z|>\alpha$. The explicit representations above will be used later to derive the large $n$ constant $E(a)$ as well as the leading asymptotics of $f^{\pm}$ in the scaling region. 

Here we should mention another important property of $f(z,\alpha)$ and $\tilde f(z,\alpha,\alpha_1)$ which can be either read from Eq.~(\ref{eq:falter}) or from the explicit representations Eq.~(\ref{eq:fplus}), Eq.~(\ref{eq:fminus}): in the $\alpha \rightarrow 1^-$ limit, away from the ``scaling region'' $|z-1|={\cal O}(1-\alpha)$, one has the following limiting forms
\begin{align}
f(z,\alpha)\bigg|_{\alpha \rightarrow 1^-} \rightarrow \frac{\pi}{2}\sqrt{-\frac{1}{z}} \ , \label{eq:limitingf} \\
\tilde f(z,\alpha,\alpha_1)\bigg|_{\alpha \rightarrow 1^-} \rightarrow \frac{\pi}{2} \sqrt{-z^{-1}}\frac{2}{\sqrt{6-z-z^{-1}}}  \ . \label{eq:limitingtildef} 
\end{align}
The above will be used later in Sec.~\ref{sec:III} as well.

Given the additive Wiener-Hopf factorization of $f$, $\tilde f$ and due to the first-order polynomial nature of $J(z)$ and $\tilde J(z)$, one obtains the {\it commutative Wiener-Hopf factorization}~\cite{commutative} for the symbol matrix $a$ 
\begin{align}
a(z)=\phi_+(z)\phi_-(z)=\phi_-(z)\phi_+(z) \ , \\ 
\phi_{\pm}(z)=\exp \bigg(J(z)f^{\pm}(z,\alpha)\bigg) \ ,
\end{align}
and similarly for the symbol matrix $\tilde a$
\begin{align}
\tilde a(z)=\tilde \phi_+(z)\tilde \phi_-(z)=\tilde \phi_-(z)\tilde \phi_+(z) \ , \\ 
\tilde \phi_{\pm}(z)=\exp \bigg(\tilde J(z)\tilde f^{\pm}(z,\alpha,\alpha_1)\bigg) \ .
\end{align}
Clearly, $\phi_{\pm}$ and $\tilde \phi_{\pm}$  are analytic in the regions $|z|<1$ (for $+$) or $|z|>\alpha$ (for $-$),  and $\phi_-$, $\tilde \phi_-$ and their inverses remain bounded as $z\rightarrow \infty$. Furthermore, at $z=0$ or $z=\infty$, $\phi_{\pm}(z)$ are upper or lower triangle matrices with diagonal elements all equals to $1$, and $\tilde \phi_{\pm}(z)$ are also constant matrices with unit determinants. The above essentially determines the $\phi_{\pm}$ and $\tilde \phi_{\pm}$ in the left or right decompositions up to constant matrices $\phi_+ \rightarrow  \phi_+ L$, $\phi_- \rightarrow L^{-1} \phi_-$ for the $+-$ left decomposition, and   $\phi_+ \rightarrow  R\phi_+ $, $\phi_- \rightarrow \phi_- R^{-1}$ for the $-+$ right decomposition. We should note that although the factors $\phi_{\pm}$ commute at the same $z$, they still do not commute at different $z$. As we will show later, this has important consequences. 

\section{Asymptotics of block determinants and scaling limits of the correlators}\label{sec:III}
Given the Wiener-Hopf factorizations, in this section we return to the correlator Eq.~(\ref{eq:corrrelator}). As known in the literature~\cite{WIDOM1974284,Basor_2007}, 
the presence of Wiener-Hopf for $a$ and $\tilde a$ with  bounded $\phi_{\pm}$, $\phi_{\pm}^{-1}$, $\tilde \phi_{\pm}$, $\tilde \phi_{\pm}^{-1}$ implies that $T(a_\eta)$, $T(a_\eta^{-1})$, $T(\tilde a_\eta)$, $T(\tilde a_\eta^{-1})$ are all invertible, where $a_\eta(z)=a(\eta z)$ with $\alpha<\eta<1$. As a result, in the $n\rightarrow \infty$ limit one always has (notice that $\ln {\rm det}\ a_{\eta}(e^{i\theta})|_{-\pi \le \theta\le \pi}\equiv 0$)
\begin{align}
D_n(a) \equiv D_n(a_\eta) \rightarrow E(a)\equiv {\rm det} \  T(a_\eta)T(a_\eta^{-1}) \ne 0 \ , \\ 
D_n(\tilde a) \equiv D_n(\tilde a_\eta) \rightarrow E(\tilde a)\equiv {\rm det} \  T(\tilde a_\eta)T(\tilde a_\eta^{-1}) \ne 0 \ .
\end{align}
Notice that $E(a)$ and $E(\tilde a)$ are clearly $\eta$ independent. Moreover, the finite-$n$ corrections to the $E(a)$ and $E(\tilde a)$ can be conveniently expressed as Fredholm determinants using the BOCG-identity~\cite{borodin1999fredholm,basor2000toeplitz}. From the BOCG representations, the scaling limits of the correlators can be further obtained in the form of exponential form factor expansions. In the rest of this section, we will follow the spirit of the above discussion to investigate in detail the properties of $D_n(a)$ and $D_n(\tilde a)$. In particular, we will derive all the major results summarized at the end of Sec.~\ref{sec:summary}.

\subsection{Calculation of $E(a)$ and $E(\tilde a)$}\label{sec:E}
In this subsection, we calculate the constants $E(a)$ and $E(\tilde a)$ for the large $n$ asymptotics of the block determinants. In particular, we derive their leading $\alpha \rightarrow 1^-$ asymptotics Eq.~(\ref{eq:Eafinal}) and Eq.~(\ref{eq:Eatildefinal}).

Before moving to the calculation details, we should mention that the fact $E(a),E(\tilde a) \ne 0$ is consistent with the physical expectation that the magnetization should be non-vanishing on the $\beta/\beta^{\star}$ boundary. In fact, for the Ising model given by Eq.~(\ref{eq:Isingaction}), due to the fact that $ \beta_c<\frac{\beta+\beta^{\star}}{2}<\beta$ and $\beta^{\star}>0$, the magnetization on the $k=0$ row is bounded from below by the magnetization on the $\frac{\beta+\beta^{\star}}{2}/0$ boundary, which is just the standard boundary magnetization for a $T<T_c$ homogeneous 2D Ising model and is well known to be non-vanishing~\cite{PhysRev.162.436}.

To proceed, we first show that in a way similar to the strong Szegö's theorem~\cite{szego1952certain}
 for scalar symbols, one still has the following compact expressions for $E(a)$ and $E(\tilde a)$ in terms of scalar functions 
\begin{align}
\ln E(a)=&\sum_{k=0}^{\infty}(2k+1)f_{-k}f_{k+1} \ , \label{eq:Ea} \\ 
\ln E(\tilde a)=& \sum_{k=0}^{\infty}(2k+1)\tilde f_{-k}\tilde f_{k+1}\nonumber \\ 
&+\frac{2\alpha_1}{(1-\alpha_1)^2}\sum_{k=0}^{\infty}\tilde f_{-k}\bigg((2k+1)\tilde f_{k+1}-k\tilde f_{k}-(k+1)\tilde f_{k+2}\bigg) \ . \label{eq:tildeEa}
\end{align}
To derive the above,  we need to use the formula~\cite{WIDOM1974284} for the derivative of $\ln E(t)=\ln E(a(t,z))$  where the symbol matrix $a(t,z)$ depends smoothly on certain parameters $t$ 
\begin{align}
&\frac{d\ln E(t)}{dt}=-\frac{1}{2\pi i}\oint_{C_\eta} dz {\rm Tr}\frac{da}{dt}\left(\frac{dU^+}{dz}U^--\frac{dV^-}{dz}V^+\right) \ , \label{eq:widom}\\ 
& a^{-1}=U^+U^-=V^-V^+ \ . 
\end{align}
Notice that the above is invariant under the transformation $a(z)\rightarrow a(\eta' z)$, as far as the integration path lies in the analyticity domain $\alpha<|z|<1$ for the un-rescaled $z$. 
Now, for commutative Wiener-Hopf, one has $a=e^{(f^++f^-)J}$, $U^{\pm}=V^{\pm}=e^{-f^{\pm}J}$. Introducing the parameter $t$ as $f^{\pm} \rightarrow t f^{\pm}$,  one has
\begin{align}
    &\frac{da}{dt}=(f^{+}+f^{-})Je^{t(f^++f^-)J} \ , \\
    &\frac{dU^{\pm}}{dz}=-tJ\left(\frac{df^{\pm}}{dz}+\frac{f^{\pm}}{2z}\right)e^{-t f^{\pm} J}-\frac{\sinh tf^{\pm}\sqrt{z}}{\sqrt{z}}\left(\frac{dJ}{dz}-\frac{J}{2z}\right) \ .
\end{align}
To simplify the above, one needs the formula
\begin{align}\label{eq:cancel}
{\rm Tr} Je^{t(f^++f^-)J}\left(\frac{dJ}{dz}-\frac{J}{2z}\right) e^{-tf^{\pm }J}\equiv 0 \ .
\end{align}
In this manner, the $\frac{dJ}{dz}-\frac{J}{2z}$ terms all drop, and after simplification one obtains
\begin{align}
\frac{d\ln E(t)}{dt}=\frac{t}{\pi i}\oint_{C_\eta} dz\bigg(2z\frac{\partial f^+}{\partial z}f^-+f^+f^-\bigg) \ .
\end{align}
Now, for $t=0$, clearly one has $E(0)=1$, thus one obtains by integrating $t$ from $0$ to $1$
\begin{align}
\ln E(a)=\frac{1}{2\pi i}\oint_{C_\eta} dz\bigg(2z\frac{\partial f^+}{\partial z}f^-+f^+f^-\bigg) \ .
\end{align}
In terms of the moments $f_i=\oint_{C_\eta} \frac{dz}{2\pi i z}z^i f(z,\alpha)$, one has
\begin{align}
\ln E(a)=\sum_{k=0}^{\infty}(2k+1)f_{-k}f_{k+1} \ .
\end{align}
The above should be compared with the strong Szegö's theorem $\sum_{k=1}^{\infty}kf_{-k}f_{k}$ for scalar symbols. 

The case of $E(\tilde a)$ can be handled similarly. Denoting $\Delta(z)=\frac{1}{(1-\alpha_1)^2}(z-\alpha_1)(1-\alpha_1z)$, one still has the crucial formula
\begin{align}\label{eq:cancel1}
{\rm Tr} \tilde Je^{t(\tilde f^++\tilde f^-)\tilde J}\bigg(\frac{d\tilde J}{dz}-\frac{1}{2\Delta(z)}\frac{d\Delta(z)}{dz}\tilde J\bigg)e^{-t\tilde f^{\pm }\tilde J}\equiv 0 \ .
\end{align}
Given the above, one proceeds as the case of $a$ to obtain
\begin{align}
&\ln E(\tilde a)=\frac{1}{2\pi i(1-\alpha_1)^2}\oint_{C_\eta} dz (z-\alpha_1)(1-\alpha_1z)\bigg(\frac{\partial \tilde f^+}{\partial z}\tilde f^--\frac{\partial \tilde f^-}{\partial z}\tilde f^+\bigg) \nonumber \\ 
&=\sum_{k=0}^{\infty}(2k+1)\tilde f_{-k}\tilde f_{k+1}+\frac{2\alpha_1}{(1-\alpha_1)^2}\sum_{k=0}^{\infty}\tilde f_{-k}\bigg((2k+1)\tilde f_{k+1}-k\tilde f_{k}-(k+1)\tilde f_{k+2}\bigg) \ .  \label{eq:tildeEaint}
\end{align}
This finishes the derivation of Eq.~(\ref{eq:Ea}) and Eq.~(\ref{eq:tildeEa}).

Given the above, let's investigate quantitative behaviors of $E(a)$ and $E(\tilde a)$, especially in the $\alpha \rightarrow 1^-$ limit. We first consider $E(a)$.  Using the representations Eq.~(\ref{eq:fkminus}) and Eq.~(\ref{eq:fkplus}) one can derive a convenient integral representation for $\ln E(a)$ which is valid for generic $0<\alpha<1$. Denoting 
\begin{align}
{\cal G}(x,\alpha)=\arctan\left(\frac{(1-\alpha)(1+x)}{2\sqrt{(1-\alpha x)(\alpha-x)}}\right) \ ,
\end{align}
the integral representation reads
\begin{align}
\ln E(a)=&\frac{1}{\pi^2}\int_{0}^{\alpha} dx \int_{0}^{\alpha}dy\frac{1}{\sqrt{xy}}\frac{1+xy}{(1-xy)^2} {\cal G}(x,\alpha){\cal G}(y,\alpha)-\frac{2}{\pi}\int_{0}^{\alpha} dx \frac{\sqrt{\alpha}{\cal G}(x,\alpha)}{\sqrt{x}(1-\alpha x)} \nonumber \\ 
&+\frac{2}{\pi}\int_{0}^{\alpha} dx \frac{{\cal G}(x,\alpha)}{\sqrt{x}(1-x)}-\sum_{k=0}^{\infty}\frac{\alpha^{k+\frac{1}{2}}(2-\alpha^{k+\frac{1}{2}})}{2k+1} \ . \label{eq:Eafull}
\end{align}
The above has been verified by numerical diagonalization of the Ising chains with system size $N=12$ for $0<\alpha\le 0.5$, with finite size errors consistent with ${\cal O}(\alpha^{N})$. 

To investigate further the behavior of $\ln E(a)$ in the $\alpha \rightarrow 1^-$ limit, one needs the formula
\begin{align}
\frac{1}{\pi^2}\int_{0}^{\alpha} dx \int_{0}^{\alpha}dy\frac{1}{\sqrt{xy}}\frac{1+xy}{(1-xy)^2} {\cal G}(x,\alpha){\cal G}(y,\alpha)\bigg|_{\alpha \rightarrow 1^-}\rightarrow\frac{2}{\pi^2}I_1 \ , \label{eq:asympt1}
\end{align}
and similarly 
\begin{align}
& -\frac{2}{\pi}\int_{0}^{\alpha} dx \frac{\sqrt{\alpha}{\cal G}(x,\alpha)}{\sqrt{x}(1-\alpha x)}\bigg|_{\alpha \rightarrow 1^-}\rightarrow-\frac{2}{\pi}I_2\ , \label{eq:asympt2} \\
& \frac{2}{\pi}\int_{0}^{\alpha} dx \frac{{\cal G}(x,\alpha)}{\sqrt{x}(1-x)}\bigg|_{\alpha \rightarrow 1^-}\rightarrow\frac{2}{\pi}I_3 \ , \label{eq:asympt3}
\end{align}
in which $I_1$, $I_2$ and $I_3$ are given by the following convergent integrals
\begin{align}
&I_1=\int_{1}^{\infty} \int_{1}^{\infty}dt_1dt_2 \frac{\arctan \frac{1}{\sqrt{t_1^2-1}}\arctan \frac{1}{\sqrt{t_2^2-1}}}{(t_1+t_2)^2}= 2 \pi  G-\frac{7 \zeta_3}{4}-\frac{1}{2} \pi ^2 \ln 2\ , \label{eq:I1}\\ 
&I_2=\int_{1}^{\infty} \int_{1}^{\infty}dt_1dt_2 \frac{\arctan \frac{1}{\sqrt{t_1^2-1}}} {(t_1+t_2)^2}=\frac{1}{2}\bigg(4G-\pi \ln 2\bigg) \ , \label{eq:I2} \\ 
&I_3=\int_{1}^{\infty} \int_{0}^{\infty}dt_1dt_2 \frac{\arctan \frac{1}{\sqrt{t_1^2-1}}} {(t_1+t_2)^2}=\frac{ \pi \ln 2 }{2} \ . \label{eq:I3}
\end{align}
Notice $G=\sum_{k=0}^{\infty}\frac{(-1)^k}{(2k+1)^2}$ is the Catalan's constant. 
The above can be understood in the following way: in the $\alpha \rightarrow 1^-$ limit, the ${\cal G}(x,\alpha)$ is strongly suppressed in the ``UV region'' in which $1-x={\cal O}(1)$, but remains ${\cal O}(1)$ in the scaling region $1-x={\cal O}(1-\alpha)$. The integrals in Eq.~(\ref{eq:asympt1}), Eq.~(\ref{eq:asympt2}), Eq.~(\ref{eq:asympt3}) are then dominated by the scaling region in the $\alpha \rightarrow 1^-$ limit as well, in which one has the approximations 
\begin{align}
&{\cal G}(x,\alpha)\bigg|_{x=1-(1-\alpha)t} \rightarrow \arctan \frac{1}{\sqrt{t^2-1}} \ ,  \\
&\frac{dx dy}{(1-xy)^2} \bigg|_{x=1-(1-\alpha)t_1, \ y=1-(1-\alpha)t_2}\rightarrow \frac{dt_1dt_2} {(t_1+t_2)^2} \ , \\
&\frac{dx}{1-\alpha x} \bigg|_{x=1-(1-\alpha)t}\rightarrow \frac{dt} {t+1} \ , \\
&\frac{dx}{1-x} \bigg|_{x=1-(1-\alpha)t}\rightarrow \frac{dt} {t}  \ ,
\end{align}
which leads to the appearance of $I_1$, $I_2$ and $I_3$. 
Given the above, in the $\alpha \rightarrow 1^-$ limit one finally has
\begin{align}
&\ln E(a)=\sum_{k=0}^{\infty}(2k+1)f_{-k}f_{k+1}\nonumber \\ 
&\rightarrow \frac{2}{\pi^2}I_1-\frac{2}{\pi}I_2+\frac{2}{\pi}I_3-\sum_{k=0}^{\infty}\frac{\alpha^{k+\frac{1}{2}}(2-\alpha^{k+\frac{1}{2}})}{2k+1} \nonumber \\ 
& \rightarrow \frac{1}{2}\ln(1-\alpha)-\frac{\ln 2}{2}-\frac{7\zeta_3}{2\pi^2} \ .
\end{align}
In the Appendix.~\ref{sec:constants}, an alternative derivation of the leading $\alpha \rightarrow 1^-$ asymptotics of $\ln E(a)$ is provided and agrees with the result derived here. 

We then move to $E(\tilde a)$ and derive its leading asymptotics in the $\alpha \rightarrow 1^-$ limit. We start from Eq.~(\ref{eq:tildeEaint}).  In the $\alpha \rightarrow 1^-$ limit , the scaling region only contributes to the first term in Eq.~(\ref{eq:tildeEaint}), due to the fact that in the second term, the three terms proportional to $k$ cancel among themselves. Introducing the function $g(z,\alpha,\alpha_1)=\tilde f(z,\alpha,\alpha_1)-f(z,\alpha)$, one has
\begin{align}
\ln E(\tilde a)-\ln E(a)=\sum_{k=0}^{\infty}(2k+1)\left(g_{-k}f_{k+1}+f_{-k}g_{k+1}+g_{-k}g_{k+1}\right) \nonumber \\ 
+\frac{2\alpha_1}{(1-\alpha_1)^2}\sum_{k=0}^{\infty}\tilde f_{-k}\bigg((2k+1)\tilde f_{k+1}-k\tilde f_{k}-(k+1)\tilde f_{k+2}\bigg) \ .
\end{align}
In the $\alpha \rightarrow 1$ limit, the above can be evaluated using the limiting symbols {\it away from the scaling region} in Eq.~(\ref{eq:limitingf}), Eq.~(\ref{eq:limitingtildef}) and
\begin{align}
g(z,\alpha,\alpha_1)\bigg|_{\alpha \rightarrow 1^-}\rightarrow  \frac{\pi}{2} \sqrt{-z^{-1}} \bigg(\frac{2}{\sqrt{6-z-z^{-1}}}-1\bigg) \ ,
\end{align}
due to the fact that $g_k={\cal O}(\frac{1}{k^2})$ in this limit. Expressing all the Fourier components as integrals along the deformed paths inside or outside the unite circles, the first sum reduces to ($\alpha_1=\frac{1}{3+2\sqrt{2}}$)
\begin{align}
&\sum_{k=0}^{\infty}(2k+1)\left(g_{-k}f_{k+1}+f_{-k}g_{k+1}+g_{-k}g_{k+1}\right)
\nonumber \\ 
&= \int_{\alpha_1}^1 dx\int_{\alpha_1}^1dy\frac{1+xy}{(1-xy)^2}\bigg(\frac{1}{4\sqrt{xy}}-\frac{1}{\sqrt{(6x-1-x^2)(6y-1-y^2)}}\bigg)\nonumber \\ 
&+\frac{1}{4}\int_{0}^{\alpha_1} dx\int_{0}^{\alpha_1}dy\frac{1+xy}{(1-xy)^2\sqrt{xy}}+\frac{1}{2}\int_{0}^{\alpha_1} dx\int_{\alpha_1}^{1}dy\frac{1+xy}{(1-xy)^2\sqrt{xy}}\nonumber \\ 
&=\frac{1}{8}+\frac{1}{2}\ln \frac{1+\alpha_1}{1-\alpha_1} \ .
\end{align}
Similarly, the second sum reduces to
\begin{align}
&\frac{1}{2}\sum_{k=0}^{\infty}k \tilde f_{-k}\left(2\tilde f_{k+1}-\tilde f_{k}-\tilde f_{k+2}\right)+\frac{1}{2}\sum_{k=0}^{\infty}\tilde f_{-k}\left(\tilde f_{k+1}-\tilde f_{k+2}\right) \nonumber \\ 
&= -\frac{1}{2}\int_{\alpha_1}^1 dx\int_{\alpha_1}^1dy\frac{(1-x)(1-y)}{(1-xy)^2\sqrt{(6x-1-x^2)(6y-1-y^2)}}= -\frac{1}{8} \ .
\end{align}
Notice that all the integral formulas above have been checked numerically. Combining the above, one has
\begin{align}
\ln E(\tilde a)-\ln E(a)=\frac{1}{2}\ln \frac{1+\alpha_1}{1-\alpha_1} \approx 0.173287 \ .
\end{align}
It is interesting to notice that this difference remains the same as the homogeneous models.

To summarize, in the $\alpha \rightarrow 1^-$ limit one has the following leading asymptotics
\begin{align}
&E(a)\rightarrow \sqrt{\frac{1-\alpha}{2}}e^{-\frac{7 \zeta_3}{2 \pi ^2}} \ , \\
&E(\tilde a)\rightarrow \sqrt{\frac{1+\alpha_1}{1-\alpha_1}}\sqrt{\frac{1-\alpha}{2}}e^{-\frac{7 \zeta_3}{2 \pi ^2}} \ . 
\end{align}
The above should be compared with the $\sqrt{2(1-\alpha)}$ of the homogeneous Ising chain and the $\sqrt{\frac{1+\alpha_1}{1-\alpha_1}}\sqrt{2(1-\alpha)}$ of the homogeneous Ising model. Notice that the $E(a)$ and $E(\tilde a)$ still scale as $\sqrt{1-\alpha}$, consistent with the scaling dimension $\frac{1}{4}$ of the spin field.

\subsection{Scaling limits of the correlators}\label{sec:f}
Now, after discussing in details the $E(a)$ and $E(\tilde a)$, in this subsection we move back to the full correlators and investigate their scaling limits. For finite $n$ one need the crucial BOCG-identity which represents the finite-$n$ corrections as Fredholm determinants ~\cite{borodin1999fredholm,basor2000toeplitz}
\begin{align}\label{eq:BOCG}
D_n(a)=E(a){\rm det} \ (1-{\cal K}_\eta) \ ,
\end{align}
where ${\cal K}_\eta$ is an operator acting on the $l^2(\{n,n+1,..\}\otimes R^2)$ with matrix elements ($\alpha<\eta<1$)
\begin{align}\label{eq:kernel}
&{\cal K}_\eta(i,j)=\eta^{-i}K(i,j)\eta^j \ , \\ 
&K(i,j)=\sum_{k=1}^{\infty} \left(\frac{\phi_+}{\phi_-}\right)_{i+k}\left(\frac{\phi_-}{\phi_+}\right)_{-j-k} \ ,
\end{align}
which are well defined due to the fact that $\phi_+$ and $\phi_-$ commute for any given $z$. The same holds for the $D_n(\tilde a)$ for which the overall constant is defined with $\tilde a$ and $\tilde a^{-1}$, and $K$ is replaced by $\tilde K$ defined with $\tilde \phi_{\pm}$.  

Notice that for more general left and right decompositions with $L$ and $R$, the kernel needs to be expressed in a way that distinguishes the left and right decompositions, but the determinant remains the same. Also notice that the $\eta$ dependency in ${\cal K}_\eta$ is simply to guarantee the boundness of the operator ${\cal K}_\eta$. Neither the determinant ${\rm det}(1-{\cal K}_\eta)$ nor the traces ${\rm Tr}({\cal K}_\eta^l)$ ($l\ge 1$) depend on $\eta$. In fact, the matrix elements $K(i,j)$ decay at large $i$ at the exponential speed $\alpha^i$, implying ${\rm Tr}(K^l)$ defined in terms of infinite sums 
\begin{align}
{\rm Tr}(K^l)\equiv \sum_{i_1, i_2,..i_l=n}^{\infty}{\rm Tr}\left(K(i_1,i_2)K(i_2,i_3)..K(i_l,i_1) \right) \ , 
\end{align}
are all finite. Moreover, one always has ${\rm Tr}(K^l) \equiv {\rm Tr}({\cal K}_\eta^l)$. Thus, the ${\rm Tr}({\cal K}_\eta^l)$-based exponential form factor expansion~\cite{Wu:1975mw,Lyberg_2007} can be performed based on ${\rm Tr}(K^l)$ in a manifestly $\eta$-independent manner.

We now investigate the scaling limit of the correlator, defined as $n=r(1-\alpha)^{-1}$ with $r$ fixed, while $\alpha \rightarrow 1^-$. As the cases of homogeneous Ising models, one expects the overall factors $E(a)$ and $E(\tilde a)$ contain all the ``UV singularities'' in the scaling limit, while the ${\rm det}(1-{\cal K}_\eta)$, ${\rm det}(1-\tilde {\cal K}_\eta)$ should allow scaling limits at the level of the exponential form factor expansion in terms of ${\rm tr} (K^l)$ and  ${\rm tr} (\tilde K^l)$. Since the Fredholm operators $K$
and $\tilde K$ are defined in terms of $\phi^{\pm}$ and $\tilde \phi^{\pm}$ which involve the scalar functions $f^{\pm}(z,\alpha)$ and $\tilde f^{\pm}(z,\alpha,\alpha_1)$, one must understand the behavior of these two scalar functions in the scaling region in order to proceed. In parallel to the scaling limit in the coordinate space, it is convenient to introduce the parameterization $z=e^{i(1-\alpha)p}$ where the $p$ plays the role of the ``momentum'' in the scaling region $|p|={\cal O}(1)$. Clearly, the upper and lower half-planes in $p$ correspond to the $|z|<1$ and $|z|>1$ regions.  

To present the result, it is convenient to introduce the functions 
\begin{align}
C(p)=1+\frac{1}{\sqrt{p^2+1}} \ , \\ 
C(p)=C_+(p)C_-(p) \ .
\end{align}
Here $C_+$ is analytic in the upper half-plane and $C_-$ is analytic in the lower half-plane. They are given by 
\begin{align}
\ln C_{\pm }(p)=\mp\int_{-\infty}^{\infty} \frac{dp'}{2\pi i}\frac{1}{p-p'\pm i0}\ln \bigg(1+\frac{1}{\sqrt{(p')^2+1}}\bigg) \ .
\end{align}
Now, to simplify the expression for $C_{+}(p)$, we notice that for $\Im p>0$ the $p'$ can be deformed to the lower half-plane to obtain
\begin{align}
\ln C_+(p)&=-\frac{1}{2\pi}\int_{-\infty}^{-1} dt \frac{1}{p-it}{\rm Disc}\ln \bigg(1+\frac{1}{\sqrt{(it)^2+1}}\bigg)=\frac{1}{\pi}\int_1^{\infty} dt \frac{\arctan \frac{1}{\sqrt{t^2-1}}}{-ip+t} \ .
\end{align}
Similarly, for the $C_-$ one simply flips $p \rightarrow -p$. They are all finite and bounded functions in the corresponding half-planes. In fact, in the large $p$ limits, one has the Mellin's representation
\begin{align}
&\ln C_{\pm}(p)=-\frac{1}{2\pi i}\int_{c-i\infty}^{c+i\infty} ds {\cal M}(s)(\mp ip)^{-s} \ , 
 \ 0<c<1 \ , \\ 
&{\cal M}(s)=-\frac{\sqrt{\pi } \Gamma \left(\frac{s}{2}+1\right)}{2s^2 \Gamma \left(\frac{s+1}{2}\right)\cos^2 \frac{\pi s}{2}}+\frac{\pi }{2 s\sin \pi s} \ .
\end{align}
Now, there is a series of double-poles at $s=2k+1$, $k\ge 0$, which leads to $\frac{1}{p^{2k+1}}\ln p$ asymptotics in the large $p$ limit, and another series of single-poles at $s=2k$, $k\ge 1$, which leads to $\frac{1}{p^{2k}}$ asymptotics. In particular, this implies that the $C_{\pm}(p)$ are bounded in the upper and lower half-planes and approach $1$ in the large $p$ limits. 

Given the above, one can state the results of the leading asymptotics of $f^{\pm}$ in the scaling region:
\begin{align}
&f^{+}(e^{i(1-\alpha)p},\alpha)\rightarrow \ln C_+(p)+\frac{1}{2}\ln (1-ip)-\ln(-ip)-\frac{1}{2}\ln(1-\alpha)+A  \ , \label{eq:asymfplus}\\ 
&f^{-}(e^{i(1-\alpha)p},\alpha)\rightarrow \ln C_-(p)+\frac{1}{2}\ln (1+ip)+\frac{1}{2}\ln(1-\alpha)-A \ , \label{eq:asymfminus}
\end{align}
where
\begin{align}
A=\ln 2  \ , 
\end{align}
is a {\it non-universal} constant due to short distance contributions,  and the remainder terms are bounded by $\sqrt{1-\alpha}$ in the scaling region. Notice that as expected, the $f^+$ is analytic in the upper half-plane, while $f^-$ is analytic in the region $\Im (p)<1$. The asymptotics Eq.~(\ref{eq:asymfplus}), Eq.~(\ref{eq:asymfminus}) holds also for the $\tilde f^{\pm}$ with a different constant  $\tilde A$ 
\begin{align}
\tilde A=A-\frac{1}{4}\int_{-\pi}^{\pi} d\theta\frac{\sqrt{-e^{-i\theta}} \left(\sqrt{6-2 \cos \theta}-2\right)e^{i\theta}}{\left(-1+e^{i \theta}\right) \sqrt{6-2 \cos \theta}} \equiv \frac{3}{4}\ln 2  \ .
\end{align}
To obtain this result, we have used the fact that $\beta_c=\frac{1}{2}\ln(1+\sqrt{2})$ and $\alpha_1=\frac{1}{3+2\sqrt{2}}$ at $\beta=\beta_c$. From the above, one has the important relation ${\color{red} e^{4A}=2e^{4\tilde A}}$ which will be used later. 

Since the constants $A$ and $\tilde A$ are crucial, here we explain how they can be derived.  The simplest way is to use the explicit representations in Eq.~(\ref{eq:fplus}) and Eq.~(\ref{eq:fminus}).
In the scaling region $z\approx 1+ i(1-\alpha)p$, similar to the Eq.~(\ref{eq:Eafull}) for $\ln E(a)$, the integrals involving the $\arctan$ functions are clearly dominated by the scaling region $x \approx 1-(1-\alpha)t$ and converge to $\ln C_{\pm}(p)$, while the sums over $k$ can be explicitly performed and expanded to obtain the remaining terms, including the constant $A=\ln 2$.  For $\tilde f$, one can write $\tilde f=f+(\tilde f-f)$ and notice that the scaling region in $\tilde f-f$ has been suppressed. In particular, in the $\alpha \rightarrow 1^-$ limit one has
\begin{align}
&-\frac{1}{2\pi i}\oint_{C_{1^-}} \frac{dz'}{1-z'}\left(\tilde f(z',\alpha,\alpha_1)-f(z',\alpha)\right)\nonumber \\ 
&\rightarrow -\frac{1}{4}\int_{-\pi}^{\pi} d\theta\frac{\sqrt{-e^{-i\theta}} \left(\sqrt{6-2 \cos \theta}-2\right)e^{i\theta}}{\left(-1+e^{i \theta}\right) \sqrt{6-2 \cos \theta}} \equiv -\frac{1}{4}\ln 2  \ ,
\end{align}
which is nothing but the difference $\tilde A-A$ between the constants. Alternatively, similar to the calculation in Appendix.~\ref{sec:constants}, the constants $A$ and $\tilde A$ can also be derived by introducing the parameterization $z'=e^{i(1-\alpha)(i\epsilon+p')}$ and splitting the $dp'$ integrals into the scaling region $|p'|<\mu$ and the ``UV region'' $\mu<|p'|<\frac{\pi}{1-\alpha}$ with $1 \ll \mu \ll \frac{1}{1-\alpha}$. Then, in the scaling region one approximates using the scaling limits of the integrands, while in the UV part one approximates by setting $z=1$ and using the limiting function $f\rightarrow \frac{\pi}{2}\sqrt{-\frac{1}{z}}$ away from the scaling region. It is not hard to show that at the leading power, the $\ln \mu$ dependencies from the two regions cancel, and the constants $A$, $\tilde A$ obtained this way agree with the first method based on the explicit representations Eq.~(\ref{eq:fplus}) and Eq.~(\ref{eq:fminus}). 

Now, after investigating in detail the Wiener-Hopf factors, one moves back to the Fredholm determinant ${\rm det}(1-{\cal K}_\eta)$. By shifting the contours of the $\left(\frac{\phi_+}{\phi_-}\right)_{i+k}$ and $\left(\frac{\phi_-}{\phi_+}\right)_{-j-k}$ factors (as Fourier components, they are defined through integrals) inside or outside the region $\alpha<|z|<1$ and picking up the singularities, one obtains the following general expression for ${\rm Tr}(K)$
\begin{align} \label{eq:trace}
&{\rm Tr}(K)\equiv \sum_{i=n}^{\infty}{\rm Tr}(K(i,i))\nonumber \\ 
=&\frac{i^2(1-\alpha)^2}{2(2\pi)^2}\int_{\frac{\ln \alpha}{\alpha-1}}^{\infty} dt_1dt_2 \frac{e^{-r(t_1+t_2)}}{\cosh(1-\alpha)(t_1+t_2)-1}{\rm Tr}\left({\rm Disc} \frac{\phi_+^2(it_1)}{a(it_1)}{\rm Disc}\frac{\phi_-^2(-it_2)}{a(-it_2)} \right) \ , \nonumber \\ 
&-\frac{i(1-\alpha)^2}{2(2\pi)}\int_{\frac{\ln \alpha}{\alpha-1}}^{\infty} dt\frac{e^{-rt}}{\cosh (1-\alpha)t-1}{\rm Tr}\left({\rm Disc}\frac{\phi_+^2(it)}{a(it)}{\rm Res}_{t_2=0^+}\frac{\phi_-^2(-it_2)}{a(-it_2)} \right)\ . 
\end{align}
Here we used $r=(1-\alpha)n$, and we have adopted the following conventions
\begin{align}
 \frac{\phi_+^2(it_1)}{a(it_1)}\equiv \frac{\phi_+^2(z)}{a(z)}\bigg|_{z=e^{-(1-\alpha)t_1}} \ , \  \ 
 \frac{\phi_-^2(-it_2)}{a(-it_2)}\equiv \frac{\phi_-^2(z)}{a(z)}\bigg|_{z=e^{(1-\alpha)t_2}} \ ,
\end{align}
and 
\begin{align}
{\rm Disc} \frac{\phi_+^2(it_1)}{a(it_1)}=\phi_+^2(z)\bigg(\frac{1}{a(z+i0)}-\frac{1}{a(z-i0)}\bigg)_{z=e^{-(1-\alpha)t_1}} \ , \\ 
{\rm Disc} \frac{\phi_-^2(-it_2)}{a(-it_2)}=\phi_-^2(z)\bigg(\frac{1}{a(z-i0)}-\frac{1}{a(z+i0)}\bigg)_{z=e^{(1-\alpha)t_2}} \ .
\end{align}
Notice that to obtain the above,  we have wrote $\frac{\phi_+}{\phi_-}=\frac{\phi_+^2}{a}$ and $\frac{\phi_-}{\phi_+}=\frac{\phi_-^2}{a}$ in the analyticity domain $\alpha<|z|<1$. Then, since $\phi_+$ and $\phi_-$ themselves are analytic inside or outside, singularities inside or outside can be read directly from $a^{-1} \equiv a_{11}{\rm 1}-a_{12}J$.  Due to the fact that $a^{-1}$ has a pole at $t_2=0^+$, the combination ${\rm Disc}\frac{\phi_+^2(it_1)}{a(it_1)}{\rm Res}\frac{\phi_-^2(-it_2)}{a(-it_2)}$ leads to the leading exponential decay. Naively, one expects that in the $\alpha \rightarrow 1^-$ limit, the $J(e^{\pm (1-\alpha)t})\rightarrow \sigma_x$ and one can simply reduce all the matrices into the form $e^{f\sigma_x}$. However, due to the presence of $\ln(1-\alpha)$ in the $f^{\pm}$, the matrix $J(e^{\pm(1-\alpha)t})$ can not be replaced by the $\sigma_x$ at the beginning, since the $\ln(1-\alpha)$ term, after exponentiation, can be amplified.

Now, to calculate the last line in Eq.~(\ref{eq:trace}), one needs the identity
\begin{align}
a^{-1}(z)=\frac{(1-\alpha)(1+z)}{(1+\alpha)(1-z)}-\frac{2\sqrt{1-\alpha z}\sqrt{1-\alpha z^{-1}}}{(1+\alpha)(1-z)}J(z) \ .
\end{align}
This implies that
\begin{align}
&\phi_+^2(z)\bigg(\frac{1}{a(z+i0)}-\frac{1}{a(z-i0)}\bigg)_{z=e^{-(1-\alpha)t_1}}=-\frac{4i\sqrt{(1-\alpha z)(\frac{\alpha }{z}-1)}}{(1+\alpha)(1-z)}e^{2f_+(z)J(z)}J(z) \ , \\ 
&-{\rm Res}_{t_2=0^+}\frac{\phi_-^2(-it_2)}{a(-it_2)}=\frac{2}{1+\alpha}e^{2f_-(1)\sigma_x}(1-\sigma_x) \ .
\end{align}
Thus one has 
\begin{align}
&-\frac{i}{2\pi }{\rm Tr}\bigg({\rm Disc}_{t\ge 1}\frac{\phi_+^2(it)}{a(it)}\times {\rm Res}_{t_2=0^+}\frac{\phi_-^2(-it_2)}{a(-it_2)}\bigg) \nonumber \\ 
&=\frac{4}{\pi(1+\alpha)^2}\frac{\sqrt{(1-\alpha z)(\frac{\alpha }{z}-1)}}{(1-z)}{\rm Tr} \left(e^{2f_+(z)J(z)}J(z)e^{2f_-(1)\sigma_x}(1-\sigma_x)\right)_{z=e^{-(1-\alpha)t}} \ .
\end{align}
 Given the above, in the scaling region $t={\cal O}(1)$ one finally has 
\begin{align}\label{eq:disc}
&-\frac{i}{2\pi t^2}{\rm Tr}\bigg({\rm Disc}_{t\ge 1}\frac{\phi_+^2(it)}{a(it)}\times {\rm Res}_{t_2=0^+}\frac{\phi_-^2(-it_2)}{a(-it_2)}\bigg)\bigg|_{\alpha \rightarrow 1^-}\nonumber \\ 
&=-\frac{1}{\pi t}\frac{\sqrt{t-1}}{\sqrt{t+1}}\frac{1}{C_+^2(it)}-{\color{red}\frac{e^{4A}}{16\pi  }}\frac{(t+1)^{\frac{3}{2}}}{t^3}\sqrt{t-1}C_+^2(it) \ .
\end{align}
To obtain the above, we have used the crucial identity
\begin{align}
&{\rm Tr}\bigg(e^{(-\ln(1-\alpha)+f_1)J(e^{-(1-\alpha)t})}J(e^{-(1-\alpha)t})e^{(\ln(1-\alpha)+f_2)\sigma_x}(1-\sigma_x)\bigg) \nonumber \\ 
& \rightarrow -2e^{-f_1-f_2}{\color{red}-\frac{t^2}{8}e^{f_1-f_2}}+{\cal O}((1-\alpha)\ln(1-\alpha)) \ ,
\end{align}
where the high-order corrections are all regular functions in $t$ and $e^{f_1}$, as well as the leading asymptotics Eq.~(\ref{eq:asymfplus}), Eq.~(\ref{eq:asymfminus}) of the $f^{\pm}$ in the scaling region which implies the following identifications
\begin{align}
&f_1=2\ln C_+(it)+\ln(1+t)-2\ln t+2A \ , \nonumber \\ 
&f_2=2\ln C_-(0)-2A \equiv \ln 2-2A \ .
\end{align}
We also used the identity 
\begin{align}
 \frac{4}{\pi(1+\alpha)^2}\frac{\sqrt{(1-\alpha z)(\frac{\alpha }{z}-1)}}{(1-z)}\bigg|_{z=e^{-(1-\alpha)t}} \rightarrow \frac{1}{\pi}\frac{\sqrt{t^2-1}}{t} \ .
\end{align}
Notice the explicit dependency on the non-universal constant $A$ in Eq.~(\ref{eq:disc}). Thus, due to the ``anomaly mechanism'' which amplifies the ``would-be power corrections'' in $1-\alpha$ from $e^{(1-\alpha)t}-1$ through the exponentiation of the $\ln(1-\alpha)$ terms, non-universal short distance contributions have been promoted to the leading power. Similarly, for $\tilde a$ one proceeds in the same spirit to obtain 
\begin{align}\label{eq:disctilde}
&-\frac{i}{2\pi t^2}{\rm Tr}\bigg({\rm Disc}_{t\ge 1}\frac{\tilde \phi_+^2(it)}{\tilde a(it)}\times {\rm Res}_{t_2=0^+}\frac{\tilde \phi_-^2(-it_2)}{\tilde a(-it_2)}\bigg)\bigg|_{\alpha \rightarrow 1^-}\nonumber \\ 
&=-\frac{1}{\pi t}\frac{\sqrt{t-1}}{\sqrt{t+1}}\frac{1}{C_+^2(it)}-{\color{red}\frac{e^{4\tilde A}}{8\pi  }}\frac{(t+1)^{\frac{3}{2}}}{t^3}\sqrt{t-1}C_+^2(it) \ .
\end{align}
 To obtain this result, we have used the explicit definition of $\alpha_1=e^{-2(\beta+\beta^{\star})}$ to express $\alpha_1$ as a function of $\alpha$ in order to expand. Notice the factor-of-two difference for the coefficients of $e^{4A}$ and $e^{4\tilde A}$.

More generally, one has the following formula in the $\alpha \rightarrow 1^-$ limit
\begin{align}\label{eq:anomaly}
&{\rm Tr }\bigg(e^{(-\ln (1-\alpha)+f_1)J(e^{-(1-\alpha)t_1})}e^{(\ln (1-\alpha)+f_2)J(e^{(1-\alpha)t_2})}\bigg) \nonumber \\ 
& \rightarrow e^{f_1+f_2}+e^{-f_1-f_2}{\color{red}-\frac{1}{16}e^{f_1-f_2}(t_1+t_2)^2}+{\cal O}((1-\alpha)\ln (1-\alpha))(e^{\pm f_1}, e^{\pm f_2}) \ ,
\end{align}
where the error terms are all regular functions in $f_1$ and $f_2$. Clearly, the first two terms correspond to the naive scaling limit in which one replaces $J=\sigma^x$ at the very beginning, while the third term shown in red corresponds to the ``anomalous contribution''. From the above, it is also clear that the power corrections in $f_1$ and $f_2$ remain power corrections after the matrix exponentiation and will not be enhanced further through infinitely many logarithms. The anomalous contribution is mainly due to the non-commutativity of the polynomial matrices $J(e^{-(1-\alpha)t_1})$ and $J(e^{(1-\alpha)t_2})$. At $t_1=-t_2$, the two matrices commute, and the anomalous contributions vanish. As such, this ``anomalous'' contribution is a unique feature of block determinants. Similarly, for the Ising model's case one also has
\begin{align}\label{eq:anomalytilde}
&{\rm Tr }\bigg(e^{(-\ln (1-\alpha)+\tilde f_1)\tilde J(e^{-(1-\alpha)t_1})}e^{(\ln (1-\alpha)+\tilde f_2)\tilde J(e^{(1-\alpha)t_2})}\bigg) \nonumber \\ 
& \rightarrow e^{\tilde f_1+\tilde f_2}+e^{-\tilde f_1-\tilde f_2}{\color{red}-\frac{1}{8}e^{\tilde f_1-\tilde f_2}(t_1+t_2)^2}+{\cal O}((1-\alpha)\ln (1-\alpha))(e^{\pm \tilde f_1}, e^{\pm \tilde f_2}) \ ,
\end{align}
with the anomalous term shown in red. Notice that the anomalous terms for the two models shown in red differ by a factor of two.

Now, given the crucial formulas Eq.~(\ref{eq:BOCG}), Eq.~(\ref{eq:trace}), Eq.~(\ref{eq:disc}) and due to the fact that the none scaling region $t\gg 1$ is exponentially suppressed by the $e^{-tr}$ factors, one obtains by taking the scaling limit in the last line of Eq.~(\ref{eq:trace}), the leading large $r$ asymptotics of the scaling function
\begin{align}
F^2_H(r) \equiv \lim_{\alpha \rightarrow 1^{-}}\frac{D_n(a)}{E(a)}\bigg|_{n=r(1-\alpha)^{-1}}  \ , 
\end{align}
as
\begin{align}
F^2_H(r)\rightarrow &1+\frac{1}{\pi}\int_1^{\infty} \frac{dt}{t}\frac{\sqrt{t-1}}{\sqrt{t+1}}\frac{1}{C_+^2(it)}e^{-tr}\nonumber \\ 
&+{\color{red}\frac{e^{4A}}{16\pi}}\int_{1}^{\infty}\frac{dt}{t^3}(t+1)^{\frac{3}{2}}\sqrt{t-1}C_+^2(it)e^{-tr}+ {\cal O}(e^{-2r}) \ .
\end{align}
Notice we have used the standard expansion ${\rm det} \ (1-{\cal K}_\eta)=e^{-\sum_{n=1}^{\infty}\frac{1}{n}{\rm tr}({\cal K}^n_\eta)}$ and the fact that ${\rm tr}({\cal K}^l_\eta)\equiv {\rm tr}(K^l)$. 
Similarly, for the Ising model's version, one obtains from Eq.~(\ref{eq:disctilde}), the leading large $r$ asymptotics of the scaling function
\begin{align}
F^2_\beta(r) \equiv \lim_{\alpha \rightarrow 1^{-}}\frac{D_n(\tilde a)}{E(\tilde a)}\bigg|_{n=r(1-\alpha)^{-1}} \ , 
\end{align}
as
\begin{align}
F_{\beta}^2(r)\rightarrow &1+\frac{1}{\pi}\int_1^{\infty} \frac{dt}{t}\frac{\sqrt{t-1}}{\sqrt{t+1}}\frac{1}{C_+^2(it)}e^{-tr}\nonumber \\ 
&+{\color{red}\frac{e^{4\tilde A}}{8\pi}}\int_{1}^{\infty}\frac{dt}{t^3}(t+1)^{\frac{3}{2}}\sqrt{t-1}C_+^2(it)e^{-tr}+ {\cal O}(e^{-2r}) \ .
\end{align}
One must notice that despite the explicit presence of the non-universal constants $A$ and $\tilde A$ in the scaling functions, since ${\color{red} e^{4A-4\tilde A}=e^{4\ln 2-3\ln 2} \equiv 2}$, the scaling functions in the two formulations are still equal, at least to the order we reached. By expanding the integrands around $t=1$, one obtains the leading asymptotics
\begin{align}
&F_H^2(r)-1 \rightarrow \frac{1}{2\sqrt{2\pi}}\frac {e^{-r}}{r^{\frac{3}{2}}}\bigg(\frac{1}{C_+^2(i)}+4C_+^2(i)\bigg)\bigg(1+{\cal O} \left(\frac{1}{r}\right)\bigg) \ , \\ 
&F_\beta^2(r)-1 \rightarrow \frac{1}{2\sqrt{2\pi}}\frac {e^{-r}}{r^{\frac{3}{2}}}\bigg(\frac{1}{C_+^2(i)}+4C_+^2(i)\bigg)\bigg(1+{\cal O} \left(\frac{1}{r}\right)\bigg) \ .
\end{align}
Furthermore, the number $C_+(i)$ is
\begin{align}
\ln C_+(i)=\frac{1}{\pi}\int_1^{\infty} dt \frac{\arctan \frac{1}{\sqrt{t^2-1}}}{t+1} \equiv \frac{1}{2\pi}\bigg(4G-\pi \ln 2\bigg) \ ,
\end{align}
where $G=\sum_{k=0}^{\infty}\frac{(-1)^k}{(2k+1)^2}$ is the Catalan's constant.  Notice that the $\frac{e^{-r}}{r^{\frac{3}{2}}}$ form of the leading exponential tail is qualitatively the same as on the $\beta/0$ boundaries with $\beta>\beta_c$~\cite{PhysRev.149.380} and should hold for all such boundaries between $\beta_1>\beta_c$ and $\beta_2<\beta_c$.

More generally, the scaling function allow the following exponential form factor expansions
\begin{align}
    F_H^2(r)=\exp \bigg(\sum_{k=1}^{\infty}\frac{1}{\pi^k}f_k(r)\bigg) \ , \\ 
    F_\beta^2(r)=\exp \bigg(\sum_{k=1}^{\infty}\frac{1}{\pi^k}\tilde f_k(r)\bigg) \ .
\end{align}
Here the ``one-particle'' form factor reads
\begin{align}
f_1(r)=\tilde f_1(r)=\int_1^{\infty} dt\frac{\sqrt{t^2-1} e^{-tr}}{t^2}\bigg(\frac{t}{g(t)}+\frac{g(t)}{t}\bigg) \ ,
\end{align}
where we have defined
\begin{align}
g(t)=C_+^2(it)(1+t)\equiv (1+t)\exp \bigg(\frac{2}{\pi}\int_1^{\infty} du \frac{\arctan \frac{1}{\sqrt{u^2-1}}}{t+u}\bigg)  \ .
\end{align}
To derive the ``two-particle'' contribution, one needs both the ${\rm tr}(K)$, ${\rm tr}(\tilde K)$ and the ${\rm tr }(K^2)$, ${\rm tr}(\tilde K^2)$. For the former, one needs the identities Eq.~(\ref{eq:anomaly}), Eq.~(\ref{eq:anomalytilde}). For the ${\rm tr }(K^2)$, ${\rm tr}(\tilde K^2)$, one needs the following identities for the $\alpha \rightarrow 1^-$ limits
    \begin{align} \label{eq:traceks}
    &{\rm Tr }\bigg(e^{(-\ln (1-\alpha)+f_1)J(e^{-(1-\alpha)t_1})}e^{(\ln (1-\alpha)+f_2)J(e^{(1-\alpha)t_2})}\nonumber \\ 
    &\times e^{(-\ln (1-\alpha)+f_3)J(e^{-(1-\alpha)t_3})}e^{(\ln (1-\alpha)+f_4)J(e^{(1-\alpha)t_4})}\bigg) \nonumber \\ 
    & \rightarrow e^{f_{1234}}+e^{-f_{1234}}{\color{red}-\frac{e^{f_{134}-f_2}}{16}t_{12}t_{23}-\frac{e^{f_{123}-f_4}}{16}t_{34}t_{41}-\frac{e^{f_1-f_{234}}}{16}t_{41}t_{12}-\frac{e^{f_3-f_{124}}}{16}t_{23}t_{34}} \nonumber \\ 
    &{\color{red}+\frac{e^{f_{13}-f_{24}}}{256}t_{12}t_{23}t_{34}t_{41}} \ ,
    \end{align}
    and 
\begin{align}\label{eq:traceks1}
    &{\rm Tr }\bigg(e^{(-\ln (1-\alpha)+f_1)\tilde J(e^{-(1-\alpha)t_1})}e^{(\ln (1-\alpha)+f_2)\tilde J(e^{(1-\alpha)t_2})}\nonumber \\ 
    &\times e^{(-\ln (1-\alpha)+f_3)\tilde J(e^{-(1-\alpha)t_3})}e^{(\ln (1-\alpha)+f_4)\tilde J(e^{(1-\alpha)t_4})}\bigg) \nonumber \\ 
    & \rightarrow e^{f_{1234}}+e^{-f_{1234}}{\color{red}-\frac{e^{f_{134}-f_2}}{8}t_{12}t_{23}-\frac{e^{f_{123}-f_4}}{8}t_{34}t_{41}-\frac{e^{f_1-f_{234}}}{8}t_{41}t_{12}-\frac{e^{f_3-f_{124}}}{8}t_{23}t_{34}} \nonumber \\ 
    &{\color{red}+\frac{e^{f_{13}-f_{24}}}{64}t_{12}t_{23}t_{34}t_{41}} \ .
    \end{align}
 Here we have adopted the notation $f_{J}=\sum_{i\in \{J\}}f_i$, $t_{J}=\sum_{i\in \{J\}}t_i$. Given the above, one proceeds as the case of $f_1(r)$ to obtain
\begin{align}
&f_2(r)+\frac{1}{2}f_1^2(r) \nonumber \\ 
&=\int_1^{\infty}\int_1^{\infty} dt_1dt_2 \frac{\sqrt{(t_1^2-1)(t_2^2-1)}e^{-t_{12}r}}{t_1t_2t_{12}^2}\bigg(\frac{t_1^2}{g(t_1)g(t_2)}+\frac{g(t_1)g(t_2)}{t_1^2}+{\color{red}\frac{t^2_{12} g(t_1)}{t_1^2g(t_2)}}\bigg) \ , \\
&\tilde f_2(r)\equiv f_2(r) \ ,
\end{align}
where $t_{12}=t_1+t_2$. In particular, the scaling functions in the two formulations remain the same up to the second order in the exponential form factor expansion. Clearly, this is due to the fact that the limiting formulas Eq.~(\ref{eq:traceks}) and Eq.~(\ref{eq:traceks1}) still relate to each other through the substitution $\frac{e^{4A}}{16\pi} \rightarrow \frac{e^{4\tilde A}}{8\pi}$, which is the identity transformation due to ${\color{red} e^{4A}=2e^{4\tilde A}}$. Notice to obtain the third order in the exponential form factor expansion, one do not need new limiting formulas since the only terms required at this order from ${\rm tr} (K^3)$ or ${\rm tr}(\tilde K^3)$ can be simply calculated as $\frac{1}{3}f_1^3$ due to the properties of the kernels $\sum_{k=1}^{\infty}\left(\frac{\phi_+^2}{a}\right)_{i+k}e^{2f_-(1)\sigma_x}(1-\sigma_x)$ and $\sum_{k=1}^{\infty}\left(\frac{\tilde \phi_+^2}{\tilde a}\right)_{i+k}e^{2\tilde f_-(1)\sigma_x}(1-\sigma_x)$. In particular, one has $\tilde f_3(r)\equiv f_3(r)$ where 
\begin{align}
&f_3(r)-\frac{1}{3}f_1^3(r)=-\int_1^{\infty}\int_1^{\infty}\int_1^{\infty} dt_1dt_2dt_3 \frac{\sqrt{(t_1^2-1)(t_2^2-1)(t_3^2-1)}e^{-t_{123}r}}{t_1^3t_3^3t_{12}t_2t_{23}} \nonumber \\ 
&\times \frac{t_1^3t_3^3+{\color{red}g^2(t_1)g^2(t_2)g^2(t_3)+g^2(t_1)g^2(t_3)t_{12}t_{23}+g^2(t_1)t_3^3t_{12}+g^2(t_3)t_1^3t_{23}}}{g(t_1)g(t_2)g(t_3)} \ .
\end{align}
Here $t_{ij}=t_{i}+t_{j}$ and $t_{123}=t_1+t_2+t_3$.
Beyond the third order, the exponential form factor expansions can all be systematically generated from ${\rm tr}(K^l)$, ${\rm tr}(\tilde K^l)$, as far as one can obtain the corresponding limiting formulas similar to Eq.~(\ref{eq:traceks}) and Eq.~(\ref{eq:traceks1}) but with $\prod_{k=1}^n e^{(-\ln(1-\alpha)+f_{2k-1})J(e^{-(1-\alpha)t_{2k-1}})}e^{(\ln(1-\alpha)+f_{2k})J(e^{(1-\alpha)t_{2k}})}$ in the trace.

\section{Summary and comments}\label{sec:conclude}
In this work, based on explicit Wiener-Hopf factorization of the symbol matrices, we managed to perform exact analysis of the correlation functions Eq.~(\ref{eq:defcorre}) and Eq.~(\ref{eq:correchain}) and study their quantitative properties in the scaling region. Before ending this work, let's make the following comments:
\begin{enumerate}
\item Due to the representations in Eq.~(\ref{eq:Isingqua}) and Eq.~(\ref{eq:correchain}), these correlators can be regarded as natural candidates for the lattice construction of the ``Sine-Gordon'' correlator (with free fermion parameters) 
\begin{align}\label{eq:sinegordon}
f^2(r)=\lim_{L\rightarrow \infty}\frac{\langle \Omega_{-m}|\sin \frac{\Phi(r)}{2}\sin \frac{\Phi(0)}{2}|\Omega_m\rangle}{\langle \Omega_{-m}|\Omega_m\rangle} \ ,
\end{align}
    between two ground states with $+m$ and $-m$ of the free-fermion masses. This roughly corresponds to the ``boundary/defect operator'' approach~\cite{Ghoshal:1993tm,Delfino:1994nr} which treats the direction orthogonal to the defect as the Euclidean time. Alternatively, if one switches the Euclidean time and space, our $\beta/\beta^{\star}$ boundary or defect can also be naively formulated in the continuum using the following Lagrangian density
    \begin{align}\label{eq:defect1}
    {\cal L}=\frac{i}{2}\bar \psi \partial \bar \psi +\frac{i}{2}\psi \bar \partial \psi-im(\theta(x)-\theta(-x))\bar \psi \psi \ .
    \end{align}
Since the Sine-Gordon theories with $\pm m$ share the common UV limit and differ only in the IR, or alternative, the mass term $-im(\theta(x)-\theta(-x))\bar \psi \psi $ is still introduced through ``relevant operators'', one may expect that the scaling limits of the spin-spin correlators on the $\beta/\beta^{\star}$ boundary calculated in this work should be safely reproduced in such continuum formulations. However, the very subtle nature of the scaling limit, especially the presence of ``anomalous terms'' against the naive scaling limit found in this work should be regarded as a warning sign to such arguments. Due to this, it is interesting to construct the spin-spin correlator in the bootstrap approach~\cite{Ghoshal:1993tm,Delfino:1994nr,Bajnok:2006ze,Bajnok:2009hp} starting from the representations Eq.~(\ref{eq:sinegordon}) or Eq.~(\ref{eq:defect1}) and compare the results with our scaling functions.

\item We must emphasize that the ``anomalous-contributions'' are due to the non commutativity of the Wiener-Hopf factors at different arguments, which can be probed at  the level of the Fredholm kernels $K$, $\tilde K$ and their traces. However, the ``one-point functions'' $E(a)$ and $E(\tilde a)$ are still controlled by the Wiener-Hopf factors and their derivatives at the same $z$, see Eq.~(\ref{eq:widom}). In principle, due to the presence of derivatives, there is still a chance to see the ``anomalous contributions'' in the one-point functions if the $e^{f^{\pm} J}$ factors fail to cancel completely, but the crucial formulas Eq.~(\ref{eq:cancel}) and Eq.~(\ref{eq:cancel1}) forbid this scenario. As a result, $E(a)$ and $E(\tilde a)$ are still controlled by the scalar functions in Eq.~(\ref{eq:Ea}) and Eq.~(\ref{eq:tildeEa}) in a way similar to scalar symbols. Due to this, they fail to see the ``anomalous contributions'' .

\item Notice that up to ${\rm tr}(K^2)$ and ${\rm tr}(\tilde K^2)$, although the $(1-\alpha)$ dependencies in the $J(e^{\pm(1-\alpha)t})$ and $\tilde J(e^{\pm(1-\alpha)t})$ can not be thrown away at the very beginning and leads to ``anomalous-contributions'' proportional to $t^2$, the $\alpha \rightarrow 1^-$ limits of the traces still exist 
and equal to each other between the two formulations. 
  At high orders, namely, for ${\rm tr}(K^l)$, ${\rm tr}(\tilde K^l)$ with $l\ge 3$, at the moment, we have neither found any counterexamples nor proved the existence of the scaling limits, but all the evidence so far strongly suggest the existence and the equality of the scaling functions in the two formulations.

Clearly, the precise forms of the small $r$ asymptotics of the scaling functions  should be regarded as one of their most important properties. At the moment, it is hard to see if the ``anomalous terms'' will be strong enough to modify the leading $r^{-\frac{1}{4}}$ rule, or just modify power corrections at orders $r^{\frac{3}{4}}$ and higher. Naively, one might favor the second scenario based on perturbative analysis of ${\rm det }(1+(T_n^{(0)})^{-1}\Delta_n)$, where $T_n^{(0)}=T_n(a)|_{\alpha=1}$ and $\Delta_n=T_n-T_n^{(0)}$, in a way similar to the homogeneous case~\cite{Wu:1975mw}. Naively taking the scaling limits at the level of matrix elements as in~\cite{Wu:1975mw}, at the power $r^{\frac{3}{4}}$ one encounters logarithmic UV divergences of the form $\int_{0}^1 du\int_{0}^1 du'\frac{f(u)g(u')}{|u-u'|}$  but not power divergence, and the UV divergences will persist to all powers. This seems to indicate that the ``anomalous terms'' might not be sufficient to modify the leading power part of the scaling function, consistent with the $\sqrt{1-\alpha}$ scaling of the $E(a)$ and $ E(\tilde a)$. Of course, it is also possible that the perturbative analysis based on the naive scaling limit will not work at all. In any case, more precise methods have to be adopted to really determine the fate of the scaling functions at small $r$. 

\item Finally, here we comment again on the differences between the $\ln(1-\alpha)$ terms for scalar and matrix factorizations. For scalar Wiener-Hopf, for example, for one of the simplest symbol
    \begin{align}
    C(z)=\sqrt{\frac{1-\alpha z^{-1}}{1-\alpha z}} \ ,
    \end{align}
    if one require that the $\ln C_-(z)$ vanishes at $z=\infty$, then in the scaling region, $\ln C_{\pm}(e^{i(1-\alpha)p})$ also contain $\ln(1-\alpha)$ terms. However, since such divergences are simply constants, one can always redefine the $\ln C_{\pm}(z)$ such that these divergences never appear by adding and subtracting. In particular, in the kernel $K$ of the Fredholm determinant 
    \begin{align}\label{eq:scalarys}
    K(i,j)=\sum_{k=1}^{\infty}\left(\frac{C_+}{C_-}\right)_{i+k}\left(\frac{C_-}{C_+}\right)_{-j-k} \ ,
    \end{align}
    such divergences always cancel for scalar symbols.
 In our matrix case, however, the scalar functions $f(z,\alpha)$ and $\tilde f(z,\alpha,\alpha_1)$ are multiplied by the polynomial matrices $J(z)$ and $\tilde J(z)$. As such, the $f^-$ and $\tilde f^-$ must vanish at $z=\infty$ in order for the $\phi_{-}$, $\tilde \phi_{-}$ to be bounded at $z=\infty$ and one loss the freedom of adding and subtracting to remove the $\ln(1-\alpha)$ in $f^{\pm}$ and $\tilde f^{\pm}$. Moreover, since the $\ln(1-\alpha)$ are multiplied by the $J(z)$, they can not be factorized out in the $K(i,j)$ and their traces as the scalar case. As demonstrated in the paper, this amplifies the short-distance contributions through the ``anomaly'' mechanism.

As such, one can read the following lesson from our example: the $\ln(1-\alpha)$ terms in matrix factorizations are harder to remove than the scalar cases. And when they appear, due to the non-commutative nature of matrices, there is a high chance that they can introduce anomalous contributions in the scaling limits. As such, for other quantities given by block determinants such as the ``entanglement entropy'' in certain fermionic models~\cite{its2006entropy,Its_2008}, the existence and universality of the ``massive scaling limits'' and their relationships to naive field theoretical descriptions in the continuum, should also be carefully investigated. 
    
\end{enumerate}

\acknowledgments
Y. L. is supported by the Priority Research Area SciMat and DigiWorlds under the program Excellence Initiative - Research University at the Jagiellonian University in Krak\'{o}w. 

\appendix

\section{Alternative derivation of the leading $\alpha \rightarrow 1^-$ asymptotics of  $E(a)$}\label{sec:constants}
This appendix provides an alternative derivative of the leading asymptotics Eq.~(\ref{eq:Eafinal}) of $E(a)$. We start from the summation formula Eq.~(\ref{eq:Ea}). We are interested in the $\alpha \rightarrow 1$ limit of the sum. For this purpose, notice that the $f_{-k}f_{k+1}$ sum can be evaluated directly in terms of the $\alpha=1$ limit of the symbol. For the $kf_{-k}f_{k+1}$ terms, both the UV region $k={\cal O}(1)$ and the scaling region $k={\cal O}(1-\alpha)^{-1}$ contribute, so one can split the summation into
\begin{align}
S=\sum_{k=0}^{\mu(1-\alpha)^{-1}}(2k+1)f_{-k}f_{k+1}+ \sum_{k=\mu(1-\alpha)^{-1}}^{\infty}(2k+1)f_{-k}f_{k+1} \ ,
\end{align}
where $1 \ll \mu(1-\alpha)^{-1}\ll (1-\alpha)^{-1}$.
For the first term, one evaluates using the $\alpha=1$ version with $f_{\rm cr}(e^{i\theta)})=\frac{\pi}{2}\sqrt{-e^{-i\theta}}$:
\begin{align}
S_{\rm UV} =-\sum_{k=0}^{\mu(1-\alpha)^{-1}}\frac{1}{2k+1}\rightarrow-\frac{1}{2} \ln \mu +\frac{1}{2}\ln (1-\alpha)-\frac{ \gamma_E }{2}-\ln 2 \ .
\end{align}
Notice the presence of $\frac{1}{2}\ln (1-\alpha)$ . 

For the second sum, one calculates using the expressions in the scaling region. The $\mu$ then becomes the lower cutoff for the $r$ integrals which contain $\ln \mu$ as UV divergences that must cancel with the $\ln \mu$ term in the $S_{\rm UV}$. To perform the sum, one needs the following expressions in the scaling region
\begin{align}
&f_{\pm k}\bigg|_{k=r(1-\alpha)^{-1}}\rightarrow (1-\alpha)\hat f(\pm r) \ ,  \nonumber \\ 
\hat f(r)=\int_{-\infty}^{\infty} \frac{dp}{2\pi} &e^{ipr}\bigg(\ln C(p)+\frac{1}{2}\ln(1+p^2)-\ln (-ip)\bigg) \ . 
\end{align}
Now, deforming the contours to the upper and lower half-planes, one has for $r>0$
\begin{align}
&f(-r)=\frac{1}{\pi}\int_{1}^\infty dt e^{-tr}\arctan \frac{1}{\sqrt{t^2-1}}-\frac{1}{2}\int_1^{\infty} dt e^{-tr}+\int_{0}^{\infty} dt e^{-tr} \ , \\ 
&f(r)=\frac{1}{\pi}\int_{1}^\infty dt e^{-tr}\arctan \frac{1}{\sqrt{t^2-1}}-\frac{1}{2}\int_1^{\infty} dt e^{-tr} \ .
\end{align}
In terms of the above, one has
\begin{align}
&S_{\rm IR}=\sum_{k=\mu(1-\alpha)^{-1}}^{\infty}(2k+1)f_{-k}f_{k+1}\nonumber \\ 
&\rightarrow \int_{\mu}^{\infty} dr 2r f(-r)f(r) 
= \frac{2}{\pi^2}I_1-\frac{2}{\pi}I_2+\frac{2}{\pi}I_3-\int_{\mu}^{\infty} dr \frac{e^{-r}}{r}\left(1-\frac{e^{-r}}{2}\right) \nonumber \\
& \rightarrow \frac{1}{2}\ln \mu+\frac{1}{2} \left(-\frac{7 \zeta_3}{\pi ^2}+\gamma_E +\ln 2\right) \ .
\end{align}
Notice that $I_1$, $I_2$ and $I_3$ are defined in Eq.~(\ref{eq:I1}), Eq.~(\ref{eq:I2}) and Eq.~(\ref{eq:I3}). Clearly, the $\ln \mu$ cancels with the one for $S_{\rm UV}$. Combining the $S_{\rm IR}$ and $S_{\rm UV}$, one obtains the final result
\begin{align}
\ln E(a)&\rightarrow \frac{1}{2}\ln(1-\alpha)-\frac{\ln2 }{2}-\frac{7 \zeta_3}{2 \pi ^2} \ , \\ 
E(a)&\rightarrow \sqrt{\frac{1-\alpha}{2}}e^{-\frac{7 \zeta_3}{2 \pi ^2}} \ .
\end{align}
As expected, all the Catalan's constants disappear in $E(a)$ and the result agrees with Eq.~(\ref{eq:Eafinal}).

\bibliographystyle{apsrev4-1} 
\bibliography{bibliography}

\end{document}